\documentclass[%
 aip,
 amsmath,amssymb,
 reprint,%
]{revtex4-1}

\usepackage{graphicx}
\usepackage{dcolumn}
\usepackage{bm}

\usepackage[utf8]{inputenc}
\usepackage[T1]{fontenc}
\usepackage{mathptmx}
\usepackage{etoolbox}
\usepackage{hyperref}
\hypersetup{
    colorlinks   = true,
    citecolor    = blue,
    linkcolor    = blue}
\usepackage{enumitem}

\makeatletter
\def\@email#1#2{%
 \endgroup
 \patchcmd{\titleblock@produce}
  {\frontmatter@RRAPformat}
  {\frontmatter@RRAPformat{\produce@RRAP{*#1\href{mailto:#2}{#2}}}\frontmatter@RRAPformat}
  {}{}
}%
\makeatother

\newcommand{\gkeyll}{\texttt{Gkeyll}}
\newcommand{\wham}{WHAM}

\renewcommand{\v}[1]{\ensuremath{\bm{#1}}} 
\newcommand{\gv}[1]{\ensuremath{\mbox{\bm$ #1 $}}} 
\newcommand{\uv}[1]{\ensuremath{\bm{\hat{#1}}}} 
\newcommand{\abs}[1]{\left| #1 \right|} 
\renewcommand{\d}[2]{\frac{d #1}{d #2}} 
\newcommand{\pd}[2]{\frac{\partial #1}{\partial #2}} 
 
\newcommand{\grad}[1]{\gv{\nabla} #1} 
\renewcommand{\div}[1]{\gv{\nabla} \cdot #1} 
\newcommand{\curl}[1]{\gv{\nabla} \times #1} 
\newcommand{\gradperp}[1]{{{\gv{\nabla}}_{\perp}} {#1}} 

\newcommand{\Zm}{\ensuremath{Z_m}}
\newcommand{\zm}{\ensuremath{z_m}}
\newcommand{\Bm}{\ensuremath{B_m}}

\newcommand{\Rm}{\ensuremath{R_m}}
\newcommand{\am}{\ensuremath{a_m}}
\newcommand{\Lm}{\ensuremath{L_m}}
\newcommand{\aw}{\ensuremath{a_w}}
\renewcommand{\ap}{\ensuremath{a_p}}
\newcommand{\alim}{\ensuremath{a_{\text{lim}}}}

\newcommand{\Bw}{\ensuremath{B_w}}
\newcommand{\Bp}{\ensuremath{B_p}}

\newcommand{\bhat}{\uv{b}}
\newcommand{\vB}{\v{B}}
\newcommand{\vBS}{\v{B^*}}
\newcommand{\BparS}{B_{\parallel}^*}
\newcommand{\vx}{\v{x}}
\newcommand{\vpar}{v_{\parallel}}
\newcommand{\vperp}{v_{\perp}}
\newcommand{\GJac}{J}
\newcommand{\kpar}{k_\parallel}
\newcommand{\kperp}{k_\perp}
\newcommand{\epsperp}{\epsilon_\perp}
\newcommand{\vparDot}{\dot{v}_{\parallel}}
\newcommand{\vxDot}{\dot{\v{x}}}
\newcommand{\zDot}{\dot{z}}
\newcommand{\rhos}{\rho_s}

\newcommand{\upare}{u_{\parallel e}}
\newcommand{\upari}{u_{\parallel i}}

\newcommand{\Tpare}{T_{\parallel e}}
\newcommand{\Tpari}{T_{\parallel i}}

\newcommand{\Tperpe}{T_{\perp e}}
\newcommand{\Tperpi}{T_{\perp i}}

\newcommand{\dz}{\text{d}z}

\newcommand{\dvpar}{\text{d}\vpar}
\newcommand{\dmu}{\text{d}\mu}
\newcommand{\dvw}{\text{d}\v{w}}
\newcommand{\dvtz}{\text{d}^3\v{z}}
\newcommand{\zmin}{-\Lz/2}
\newcommand{\zmax}{\Lz/2}
\newcommand{\vparmax}{v_{\parallel\max}}
\newcommand{\vparmin}{v_{\parallel\min}}
\newcommand{\mumax}{\mu_{\max}}

\newcommand{\Lz}{L_z}
\newcommand{\Dt}{\Delta t}
\newcommand{\Dz}{\Delta z}
\newcommand{\Dvpar}{\Delta\vpar}
\newcommand{\Dmu}{\Delta\mu}
\newcommand{\Nz}{N_z}
\newcommand{\Nvpar}{N_{\vpar}}
\newcommand{\Nmu}{N_\mu}

\newenvironment{eqnal}{\equation\aligned}{\endaligned\endequation}

\newcommand{\ignore}[1]{}  

\begin{document}

\preprint{AIP/123-QED}

\title[Continuum gyrokinetics of high-field mirrors]{Towards continuum gyrokinetic study of high-field mirrors}
\author{M. Francisquez}
\affiliation{Princeton Plasma Physics Laboratory, Princeton, NJ 08540, USA.}

\author{M. H. Rosen}
\affiliation{Department of Astrophysics Sciences, Princeton University, Princeton, NJ 08540 USA.}%

\author{N. R. Mandell}%
\affiliation{Princeton Plasma Physics Laboratory, Princeton, NJ 08540, USA.}%

\author{A. Hakim}
\affiliation{Princeton Plasma Physics Laboratory, Princeton, NJ 08540, USA.}

\author{C. B. Forest}
\affiliation{Department of Physics, University of Wisconsin-Madison, Madison, WI 53706, USA.}

\author{G. W. Hammett}
\affiliation{Princeton Plasma Physics Laboratory, Princeton, NJ 08540, USA.}


\begin{abstract}
High-temperature superconducting (HTS) magnetic mirrors under development exploit strong fields with high mirror ratio to compress loss cones and enhance confinement, and may offer cheaper, more compact fusion power plant candidates. This new class of devices could exhibit largely unexplored interchange and gradient-driven modes. Such instabilities, and methods to stabilize them, can be studied with gyrokinetics given the strong magnetization and prevalence of kinetic effects. Our focus here is to: a) determine if oft-used gyrokinetic models for open field lines produce the electron-confining (Pastukhov) electrostatic potential; b) examine and address challenges faced by gyrokinetic codes in studying HTS mirrors. We show that a one-dimensional limit of said models self-consistently develops a potential qualitatively reaching the analytical Pastukhov level. Additionally, we describe the computational challenges of studying high mirror ratios with open field line gyrokinetic solvers, and offer a force softening method to mitigate small time steps needed for time integration in colossal magnetic field gradients produced by HTS coils, providing a 19X speedup.
\end{abstract}

\maketitle

\section{Motivation \& background} \label{sec:intro}

The world's energy needs, the time scale on which clean energy sources are needed, and existing financial structures all favor fusion power plants that can be deployed quickly in the coming decades and with the lowest possible capital cost~\cite{NASS2021}. This ecosystem has incentivized R\&D of tokamaks~\cite{Creely2020} and stellarators~\cite{Volberg2022,Qian2022} that use emerging technologies and manufacturing techniques to shorten the time frame and reduce the cost of bringing fusion energy to the grid. Non-toroidal devices striving to achieve similar goals with less time and fewer resources
are also being explored. One such candidate is the axisymmetric magnetic mirror, a device consisting primarily of two coils cylindrically aligned to magnetically confine a plasma between them. Recent years have seen renewed interest in mirror-based fusion due to a number of reasons presented in more detail in previous studies (~\cite{Ryutov2011,Simonen2008,Fowler2017,Anderson2020,Endrizzi2023} and references therein), some of which we briefly review here. Amongst them is the realization of magnetohydrodynamic (MHD) stability and keV electron temperatures in one such mirror, the Gas Dynamic Trap (GDT) at the Budker Institute~\cite{Ryutov2011,Bagryansky2015}. These findings, and the arrival of new heating and superconducting technologies, also motivated proposals for a tandem mirror~\cite{Fowler2017} (earlier called an ambipolar trap~\cite{Ryutov1988}) to reach breakeven, when the fusion power released surpasses input heating power. A tandem mirror has a long linear plasma with an axial arrangement of coils (the ``central cell''), capped at both ends by ``end plug'' mirrors with higher energy density and stronger magnetic fields than those in the central cell.

Amongst the technological advances modern mirrors can employ is the emergence of high-temperature superconductors (HTS) based on REBCO tapes~\cite{Whyte2019}. HTS coils can operate at higher temperatures and current density, both desirable in fusion applications, and generate much stronger magnetic fields, which in the context of the mirror means smaller loss cones and synergistic operation with modern RF heating systems. These are some of the advantages the Wisconsin HTS Axisymmetric Mirror (\wham), now under construction at the University of Wisconsin, intends to leverage as it demonstrates the operation of an HTS end plug. 
\wham~will consist of two HTS coils $\approx$2 m apart, producing magnetic fields with a peak amplitude of $\approx$17 T; field lines extend past the coils to expander regions where they flare out.

Along this magnetic field, an electric field that confines the electrons and mitigates parallel particle and heat losses is expected to appear, as in previous mirrors. This ``ambipolar'' field results from light and collisional electrons in the mirror rapidly scattering into velocity space loss cones and attaining velocities that allow them to escape to the expander and eventually collide with the wall. The ions, however, stay behind because they are heavy and slow, and scatter much more slowly into the loss cone; hence a positive potential begins to build in the plasma. This rising potential increasingly decelerates outgoing electrons and accelerates the ions, and is arrested when a quasineutral equilibrium with equal ion and electron particle fluxes (ambipolarity) is reached. Analytical calculations of the particle loss rates, and the potential to support them, were first done by Pastukhov~\cite{Pastukhov1974} and soon after corrected by others~\cite{Cohen1978,Chernin1978}. They consisted of approximate solutions to the kinetic equation with the Landau/Rosenbluth collision operator (LRO), using a simplified bounce averaging with a step-function magnetic field model and neglecting expanders. Despite their simplifications, the resulting potential drops ($\Delta\phi$) from the center of the plasma to the location with the maximum magnetic field amplitude (the mirror throat) of $e\Delta\phi/T_e\sim 4-5$ overlapped with contemporaneous numerical calculations giving $e\Delta\phi/T_e\sim 4-6$ ($T_e$ is the electron temperature)~\cite{Killeen1976,Baldwin1977,Post1987}. Since then, more numerical and semi-analytic studies have targeted similar or related scenarios. For example, a semi-analytical study in the region surrounding one mirror coil in a rather collisional (GDT-like) regime obtained $e\Delta\phi/T_e\approx0.77$ followed by an additional potential drop of $\approx5T_e/e$ in the expander~\cite{Wetherton2021}. They also reported particle-in-cell (PIC) simulations giving $e\Delta\phi/T_e\approx0.47$ in the plasma and a potential drop of $\approx1.65T_e/e$ across the central plasma and the expander. For a similar geometry but different parameters, a more recent PIC work with electron parallel force balance obtained $e\Delta\phi/T_e\approx3.48$~\cite{Jimenez2022}. More relevant to our work is the recently computed $e\Delta\phi/T_e\approx5-6$ for \wham-relevant, more collisionless parameters using semi-analytical methods~\cite{Egedal2022}. 

The precise height and shape of the ambipolar potential depends on numerous factors such as mirror ratio, confinement scheme, plasma parameters, geometry, and heating methods, which all vary in the aforementioned studies. This variability is reflected in various measurements made in mirrors throughout the years. For example, in the TMX and TMX-U machines, $e\Delta\phi/T_e=1.5-2.5$ and $3.5$, respectively~\cite{Simonen2008}. Also, at a similar level to TMX-U, the potential drop measured with probes in the GAMMA 10 tandem mirror was $\approx(200/60)T_e/e=3.33T_e/e$~\cite{Yoshikawa2019}. Moreover, there have been many inferences of the ambipolar potential in GDT, some earlier ones~\cite{Bagryansky2016,Ivanov2013,Ivanov2017} yielding $(3-5)T_e/e$ and more recent ones~\cite{Lizunov2022} reporting $(2.6-3.1)T_e/e$. And although not discussed in this work, the presence of high-energy ``sloshing ions'' produced by oblique incidence of neutral beams can also modify the potential due to increased off-center peak densities resulting from the more distant turning points of said ions~\cite{Ryutov2011}.

In addition to developing a sufficiently strong ambipolar potential to diminish the electron parallel particle and energy transport, mirrors have to mitigate several other phenomena that can degrade confinement. There are cyclotron instabilities like the Alfv\'en ion cyclotron (AIC) and drift cyclotron loss cone (DCLC) modes, which feed off the radial density gradient or the positive gradient in velocity-space caused by ambipolar potentials and produce fluctuating fields at frequencies comparable to the cyclotron frequency. Additionally, there are other gradient-driven drift modes like electron and ion temperature-gradient modes (ETG, ITG) and trapped particle modes. These lower frequency modes produce transport across field lines and deteriorate confinement. Lastly, there is the severe interchange flute instability caused by the charge separation induced by the azimuthal $\grad{B}$ drift of ions and electrons and its resulting $E\times B$ drift that reinforces initial azimuthal perturbations. It is one of the most concerning owing to its short, microsecond time scale. Strategies like good-curvature~\cite{Ryutov2011}, kinetic~\cite{Post2001}, active feedback~\cite{Arsenin1977} and vortex~\cite{Beklemishev2010} stabilization have been developed to suppress this mode's growth. The latter uses electrical biasing between a ring limiter in the plasma and concentric rings at the far end of the expander to generate radially sheared azimuthal flows. These flows can suppress the growth of radial perturbations in the same way shear-flow stabilization does in tokamaks~\cite{Burrell1997}, or similar to how biasing-induced azimuthal flows reduce cross-field transport in LAPD~\cite{Carter2009}. In contrast to tokamaks, where reduction of ITG heat transport by flow shearing is done indirectly via a complex relationship between actuators and turbulence, a potential advantage of the mirror is that these flows may be tailored directly via end-plate biasing.



The effectiveness of biasing in controlling interchange and ITG modes and the expected magnitude of the turbulent cross-field transport produced by these and other low-frequency instabilities is not well characterized in \wham~or other future HTS mirrors. Luckily the tokamak community has spent decades maturing computational tools that can allow us to explore these phenomena through, for example, continuum~\cite{Idomura2009,gs2022,geneWeb,Candy2016} and PIC~\cite{Parker2006,Ku2016} gyrokinetic~\cite{Krommes2012} simulation. These codes may be capable of providing a first-principles estimation of the microinstability characteristics and the effects end-plate biasing has on them. Their design, however, was not meant for an HTS mirror environment where, for example, magnetic fields are open-ended and have large parallel gradients. Luckily in recent years, the \gkeyll~code~\cite{gkeyllWeb} pioneered continuum gyrokinetic modeling in open field lines, which may provide a sound foundation on which to build capabilities to perform first-principles simulations of low-frequency turbulence in HTS mirrors much in the same way the community has done for tokamaks for the last 20 years. A hint at what this may look like can be seen in \gkeyll~simulations of LAPD in which biasing was shown to reduce cross-field turbulent transport~\cite{Shi2017thesis}.

Hence, in this work, we explore two stepping-stone issues toward achieving predictive capability over HTS mirror turbulence. The first issue is whether the gyrokinetic model in the \gkeyll~code (also present in other gyrokinetic codes~\cite{Michels2021,Dorr2018}) is consistent with and capable of producing the Pastukhov ambipolar potential, an essential requirement for mirror-based fusion since electron parallel heat losses need to be curtailed. We point out that the main subject of a gyrokinetic code like \gkeyll~is short time scale turbulence, rendering it inefficient at calculating the Pastukhov potential, which evolves and saturates on a much longer time scale. A bounce-averaged code is much faster for this task, but demonstrating that a gyrokinetic model is consistent with the Pastukhov potential and qualitatively reproduces it on a shorter time scale is an important initial test. The second issue concerns the computational challenges brought about by the HTS mirror environment, to which tokamak-oriented codes would need to adapt and develop solutions to. Examples include extreme time step constraints from the colossal mirror force, and the high mirror ratio that increases resolution requirements and compute time (e.g. a mirror ratio of 20 requires increasing the velocity-space resolution by about a factor of 20 to resolve the loss cone). These topics are addressed in the following sections: section~\ref{sec:models} introduces the physics models employed in this work, section~\ref{sec:comp} describes some of the computational challenges in carrying out continuum gyrokinetic simulations of an HTS mirror, section~\ref{sec:results} presents our results, and section~\ref{sec:conclusion} summarizes this work.

\section{Equilibrium \& gyrokinetic models used} \label{sec:models}

The simulations presented in the next section are produced with the following model magnetic equilibrium and three different kinetic models, each detailed in this section.

\subsection{Magnetic equilibrium model} \label{sec:equilibrium}

\begin{figure}
\includegraphics[width=0.48\textwidth]{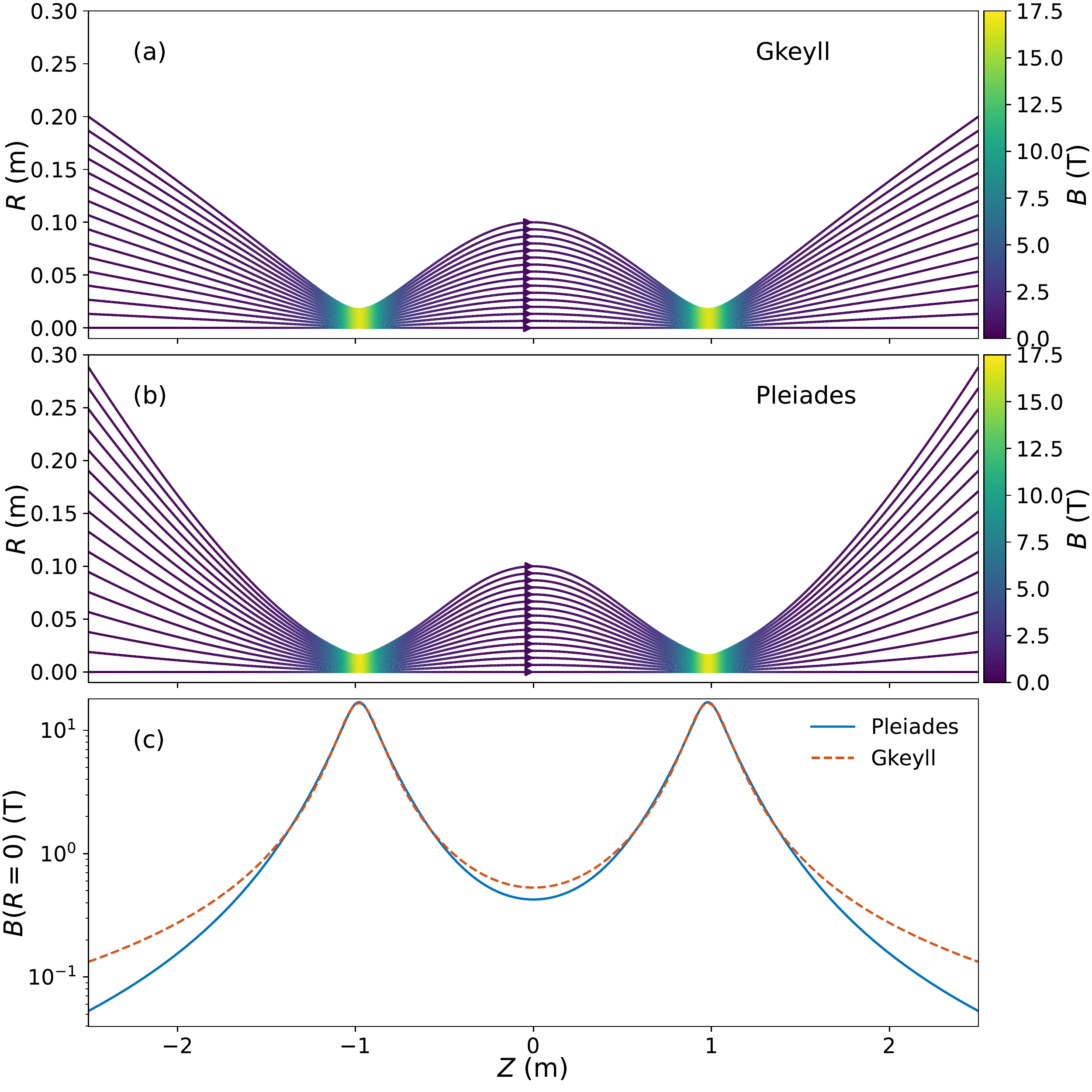}
\caption{\label{fig:eqcomp} Comparison of equilibrium magnetic field in the model used in \gkeyll~(a) and as computed by the Pleiades code for the coils in \wham~(b). The magnetic field on axis given by the model is very similar to that in the planned experiment (c).}
\end{figure}

We use a model magnetic mirror that resembles the proposed \wham~experiment, including the expanders where flux expansion and good magnetic curvature occur. This model magnetic equilibrium is composed of a radial and an axial component, $\v{B}=B_R\uv{R}+B_Z\uv{Z}$, each of which is given by a magnetic flux function $\psi(R,Z)$ as $B_R=-(1/R)\partial_Z\psi$ and $B_Z=(1/R)\partial_R\psi$. In our model, the magnetic flux function consists of a double Lorentzian: 
\begin{equation} \label{eq:psimodel}
\psi = \frac{R^2\mathcal{B}}{2\pi\gamma}\left\{\left[1+\left(\frac{Z-\Zm}{\gamma}\right)^2\right]^{-1}\hspace{-5pt}+\left[1+\left(\frac{Z+\Zm}{\gamma}\right)^2\right]^{-1}\right\},
\end{equation}
which has a good description of the field between mirror coils but is less accurate in the expanders. This simple three-parameter model was tuned to generate a magnetic field similar to that aimed for \wham. The resulting field lines and magnetic field magnitudes in the R-Z plane are shown in figure~\ref{fig:eqcomp}(a) for the parameters $\{\mathcal{B},\gamma,\Zm\}=\{6.51292,0.124904,0.98\}$, and can be compared with those obtained by the Pleiades code~\cite{Pleiades} in figure~\ref{fig:eqcomp}(b). The field is quite similar in the central region where the plasma is confined, but this Lorentzian model does not precisely capture the curvature of the field lines in the expanders. The on-axis field is shown in figure~\ref{fig:eqcomp}(c), illustrating that this model is reasonable as a first step in this study. Comparison of several equilibrium metrics for the Pleiades-computed field and the model used here is given in table~\ref{tab:eqpars}. The $\Rm=\Bm/\Bp=32$ mirror ratio for our model, although higher than the $\Rm=20$ in the Pleiades calculation, is within the $\Rm=20-50$ range that \wham~plans to explore so we did not try to improve this field further at this time. Note that this $\Rm=32$ is significantly higher than other recent mirror simulations~\cite{White2018,Jimenez2022,Wetherton2021}. A more realistic and significant discrepancy is that the magnetic field at the expander wall ($B(Z=L/2)=\Bw$) is 2.6 times larger than the Pleiades calculation and larger than what will be found on \wham~as well, which will impact things like density profiles in the expander. In the future we will perform studies with more accurate magnetic fields computed by Pleiades or other equilibrium tools.

\begin{table}
\caption{Mirror equilibrium in \wham~compared to our model} \label{tab:eqpars}
\begin{ruledtabular}
\begin{tabular}{c l c c}
parameter & description & \wham & \gkeyll \\
\hline
$L$ & Machine length & 5 m & 5 m \\
\Lm & Distance between mirror coils & 1.8 m & 1.96 m\\
\am & Mirror radius (at mirror coils) & 0.0275 m & 0.018 m \\
\ap & Plasma radius & 0.1 m & 0.1 m\\
\aw & Expander radius at wall & 0.35 m & 0.2 m\\
\Bm & Mirror field strength & 17 T & 17.016 T \\
\Bp & Plasma field strength & 0.85 T & 0.53 T \\
\Bw & Field strength at expander wall & 0.05 T & 0.1327 T
\end{tabular}
\end{ruledtabular}
\end{table}

\subsection{Gyrokinetic models} \label{sec:gkmodels}

\gkeyll~\cite{gkeyllWeb} solves a long-wavelength full-$f$ gyrokinetic model, either electrostatic or electromagnetic (i.e. with $\v{B}_\perp$ fluctuations~\cite{Mandell2020}), to determine the evolution of the distribution function of species $s$, $f_s(\v{x},\vpar,\mu,t)$, as a function of time $t$, guiding-center position $\v{x}=(x,y,z)$, velocity parallel to the background magnetic field $\vpar$ and magnetic moment $\mu=m_s\vperp^2/(2B)$. The model evolves $f_s$ by solving
\begin{equation} \label{eq:gkeq}
\pd{\BparS f_s}{t} + \div{\BparS\vxDot f_s} + \pd{}{\vpar}\left(\BparS\vparDot f_s\right) = \BparS C[f_s] + \BparS S_s
\end{equation}
in the electrostatic limit, where $\vBS=\vB+(m_s\vpar/q_s)\curl{\bhat}$ and we approximate $\BparS=\bhat\cdot\vBS\approx B$. The terms $C[f_s]$ and $S_s$ stand for collisions and sources, respectively. Collisions are modeled via a Dougherty operator~\cite{Francisquez2020}, which has been improved to better model multispecies collisions~\cite{Francisquez2022}. The phase-space characteristics are defined by 
$\vxDot=\{\vx,H\}$ and $\vparDot=\{\vpar,H\}$ in terms of the Poisson bracket
\begin{equation} \label{eq:poissonBracket}
    \{F,G\} = \frac{\vBS}{m_s\BparS}\cdot\left(\grad{F}\pd{G}{\vpar}-\grad{G}\pd{F}{\vpar}\right)-\frac{\bhat}{q_s\BparS}\times\grad{F}\cdot\grad{G},
\end{equation}
and the Hamiltonian
\begin{equation} \label{eq:hamiltonian}
    H = \frac{1}{2}m_s\vpar^2 + \mu B + q_s\phi - \frac{1}{2}m_s\abs{\v{v_E}}^2
\end{equation}
including the higher order term associated with the kinetic energy due to $E\times B$ drifts~\cite{Sugama2000,Brizard2007} of velocity $\v{v_E}=\v{E}\times\bhat/B$.

The gyrokinetic evolution described by equations~\ref{eq:gkeq}-\ref{eq:hamiltonian} is coupled to the long-wavelength Poisson equation (i.e. quasineutrality) given by
\begin{equation} \label{eq:poisson}
    -\div{\sum_s\frac{n_sm_s}{B^2}\gradperp{\phi}} = \sum_s q_sn_s
\end{equation}
where we refer to $n_s$ as the guiding center number density, defined as the zeroth velocity moment of $f_s$: $n_s=\int\dvw~\BparS f_s$ with $\dvw=(2\pi/m_s)\dvpar\,\dmu$. In general, the spatially and time-dependent density is used in the left side of equation~\ref{eq:poisson}, in contrast to the linearized polarization density or Boussinesq approximation  commonly employed in other full-$f$ gyrokinetic~\cite{Michels2021} and fluid~\cite{Tamain2016} codes. The last term in the Hamiltonian (equation~\ref{eq:hamiltonian}) is kept to guarantee energy conservation when the Boussinesq approximation is not made, demonstrated in the 1D limit in appendix~\ref{sec:energyConserv}.

The model given by equations~\ref{eq:gkeq}-\ref{eq:poisson} is cast in field-aligned coordinates, which in this case, we choose to be the equilibrium magnetic flux $\psi$, the azimuthal angle $\theta$ and the length along the field line $\ell$, i.e. our guiding center coordinates are $\v{x}=(x,y,z)=(\psi,\theta,\ell)$. \gkeyll~requires that we provide the mapping $\v{X}(\v{x})$ from field-aligned ($\v{x}$) to Cartesian ($\v{X}=(X,Y,Z)$) coordinates, from which the metric coefficients $g_{ij}=(\partial \v{X}/\partial u^i)\cdot(\partial \v{X}/\partial u^j)$ (for $u^i\in\{x,y,z\}$) and the position-space Jacobian $\GJac$ are computed using automatic differentiation routines in the SciLua package~\cite{scilua}. Given our choice of $(\psi,\theta)$ (or $(x,y)$), $R(x,Z)$ can be obtained by inverting equation~\ref{eq:psimodel}. Lastly, we obtained the mapping $Z(\v{x})$ for a given $(x,y)$ by considering the field line equation $\mathrm{d}(\v{X}(z))/\mathrm{d}z|_{\text{along}~B} = \v{B}/B$ which after dotting with \uv{Z} can be rearranged into
\begin{equation} \label{eq:zofZ}
    z(Z) = \int_0^Z\mathrm{d}s\,\frac{B(x,s)}{B_Z(s)}.
\end{equation}
In general, the solution to this equation may not be analytic, but we integrate and invert it numerically to obtain $Z(\v{x})$.

\begin{table}
\caption{Reference plasma parameters used in simulations} \label{tab:simPars}
\begin{ruledtabular}
\begin{tabular}{c l c}
parameter & description & value \\
\hline
$n_0$ & Number density & $3\times10^{19}$ m$^{-3}$ \\
$T_{e0}$ & Electron temperature & 940 eV \\
$T_{i0}$ & Ion temperature & 8.361 keV \\
$m_i$ & Ion mass & $m_{\text{deuterium}}$
\end{tabular}
\end{ruledtabular}
\end{table}

Although our ultimate goal is to study the three-dimensional dynamics of \wham, in this work, we constrain ourselves to a one-dimensional exploration of parallel dynamics along a single field line across the entire machine. Our scope is thus more comprehensive than previous ones that only focused on ``nozzles" or a single coil region~\cite{Robertson2016,Wetherton2021,Jimenez2022}. We solve a 1D limit of the above model on the field line going through $(R,Z)=(0.1\,\mathrm{m},0)$, where the evolution of the distribution function $f_s(z,\vpar,\mu,t)$ is given by
\begin{equation} \label{eq:gkeq1d}
\pd{\BparS f_s}{t} + \frac{1}{\GJac}\pd{}{z}\left(\GJac\BparS\zDot f_s\right) + \pd{}{\vpar}\left(\BparS\vparDot f_s\right) = \BparS C[f_s] + \BparS S_s,
\end{equation}
the Jacobian $J$ is now evaluated at $(\psi(R=0.1\,\mathrm{m}),\theta=0)$, and the guiding-center equations of motion are now obtained with the simpler Poisson bracket
\begin{equation} \label{eq:poissonBracket1d}
    \{F,G\} = \frac{1}{m_s\GJac\BparS}\left(\pd{F}{z}\pd{G}{\vpar}-\pd{G}{z}\pd{F}{\vpar}\right).
\end{equation}
The kinetic equation in this 1D limit is often referred to as the drift-kinetic equation; we continue to call it the gyrokinetic equation because it is a 1D limit of our higher dimensional (long-wavelength) gyrokinetic model and because, as it is shown below, we use a Poisson equation to obtain the electrostatic potential as one does in higher dimensional gyrokinetics.

As a starting point, we only consider the study of the thermal plasma with the reference parameters in table~\ref{tab:simPars} and do not directly model beam ions; this is because the use of ${\sim}25$ keV neutral beams introduces an energetic sloshing population that is more computationally expensive to account for. Although we have ideas on how we will mitigate this expense, we first want to understand the challenges associated with the thermal population and explore the Pastukhov potential formation for which the beam ions are not essential. However, we incorporate a particle and heat source at the center of the field line, mimicking the combined effect of beam ions slowing down on the thermal population, ionization particle sources and microwave heating. Such source has the following shape:
\begin{equation} \label{eq:source}
    S_s(z) = \frac{\mathcal{N}}{\left(2\pi \mathcal{T}_s/m_s\right)^{3/2}}\exp\left(-\frac{\vpar^2+2\mu B}{2\mathcal{T}_s/m_s}\right),
\end{equation}
and uses the source density rate and temperature 
\begin{eqnarray}
    &\mathcal{N} = \begin{cases}
    \max\left(\mathcal{N}_{\text{floor}},~\frac{\mathcal{N}_0}{\sqrt{2\pi\sigma^2}} e^{-z^2/(2\sigma^2)}\right) \qquad &\abs{Z}\leq Z_m \\
    \mathcal{N}_{\text{floor}} \qquad &\abs{Z}>Z_m
    \end{cases} \label{eq:sourceN} \\
    &\mathcal{T}_s = \begin{cases}
    \mathcal{T}_{s0} \qquad &\abs{z}\leq 2\sigma \\
    \mathcal{T}_{\text{floor},s} \qquad &\abs{z}>2\sigma
    \end{cases} \label{eq:sourceT}
\end{eqnarray}
with the following parameters: $\mathcal{N}_{\text{floor}}=10^{-16}$ m$^{-3}$s$^{-1}$, $\mathcal{N}_0=3.96\times10^{22}$ m$^{-3}$s$^{-1}$, $\sigma=\Zm/4$, $\mathcal{T}_{\text{floor},s}=\mathcal{T}_{s0}/8$, $\mathcal{T}_{s0}=1.25T_{s0}$. The small density source floor ($\mathcal{N}_{\text{floor}}$) is included to reduce the risk of having negative distribution functions, which the code does not forbid. Still, we try to keep it at sufficiently low levels so it does not affect the physics results. As mentioned in section~\ref{sec:intro}, sloshing ions do modify the potential profile due to their off-center peaked density, and we intend to explicitly include them in subsequent studies.

In addition to sources, the right side of the kinetic equation~\ref{eq:gkeq1d} includes a collision term that, in this study, only accounts for intra-species collisions, using a spatially constant and time-independent collision frequency~\cite{Huba2013} computed using the reference parameters in table~\ref{tab:simPars}. Future studies will include the effect of multispecies collisions with an improved Dougherty operator~\cite{Francisquez2022}. Additionally, the homogeneous collisionality causes the expanders to be more collisional than they should be since $\nu_s\propto n_s/T_s^{3/2}$ and the density drops precipitously there, but that is something we could relax in future work. Perhaps the fact that our simplified operator lacks the $1/v^3$ dependence of the LRO is of greater importance; more precisely,  the pitch-angle scattering part of the Dougherty operator corresponds to a collision frequency scaling as $1/v^2$ instead of the $1/v^3$ scaling in the LRO\footnote{The velocity drag and diffusion parts of the Dougherty operator do have a collision frequency independent of velocity, in contrasts with the LRO's $1/v^3$ dependence for equivalent terms}. Such scaling leads to somewhat larger scattering rates at high energies and different equilibrium potentials. Improving the model to incorporate the $1/v^3$ scaling and implementing the LRO are subjects of ongoing work.

Alongside the Dougherty operator, the sources in equations~\ref{eq:source}-\ref{eq:sourceT} complete the definition of the kinetic equation we evolve. Integration, however, also requires boundary (BCs) and initial (ICs) conditions. As for BCs, the field line in our domain terminates at the sheath entrance since the sheath is a non-neutral region beyond the scope of gyrokinetics. At this location, we apply a conducting sheath BC~\cite{Shi2017}, in which the potential $\phi$ is determined by the field model (below) and particles with $m_s\vpar^2/2 < -q_s\phi$ are reflected (assuming zero wall potential) while the rest are absorbed by the wall.
At velocity space boundaries, we apply zero-flux BCs, as needed for our model, to conserve moments of $f_s$. Such BCs are applied in every step from $t=0$, at which time we begin with Maxwellian ICs that have the following moments:
\begin{eqnarray} \label{eq:nTics}
    n_s(z,t=0) = A(z; n_0, n_m), \\
    T_s(z,t=0) = A(z; T_{s0}, T_{sm}), \\
    u_{\parallel s}(z,t=0) = c_{sm}\,z/z_m
\end{eqnarray}
in terms of the helper functions
\begin{eqnarray}
    &A(z; h_0, h_m) = \begin{cases}
        h_0\, L(z) \quad &\abs{Z}\leq Z_{bt}, \\
        h_0\, L(z)^5 \quad &Z_{bt} < \abs{Z}\leq Z_{m}, \\
        h_m\sqrt{B(z)/B_m} \quad &Z_{m} < \abs{Z},
    \end{cases} \label{eq:nTicsA} \\
    &L(z) = 1-\left[\left(R(z)-R_{bt}\right)/a_{\mathrm{lim}}\right]^2. \label{eq:nTicsL}
\end{eqnarray}
The function $A(z; h_0, h_m)$ is chosen such that the density increases from $(R=0.1\,\mathrm{m},Z=0)$, where we assume particles are injected at a 45-degree angle, towards the turning point of trapped orbits at $(R_{bt}=0.07\,\mathrm{m},Z_{bt}=0.47\,\mathrm{m})$, where $B(z)=2B_p$ (from conservation of kinetic energy and $\mu$). A limiter radius of $a_{\mathrm{lim}}=0.125\,\mathrm{m}$ is assumed. The profiles in the expander would preferably follow the ratio $B(z)/\Bp$ rather than its square root, but the latter was chosen to ease numerical transients earlier in the simulation; $A(z;h_0,h_m)$ in $Z_{bt} < \abs{Z}\leq Z_{m}$ is simply chosen as a polynomial connection between the other two regions. These profiles give initial densities and temperatures at the mirror throats given by $n_m$ and $T_{sm}$, respectively, from which one can define the parameter $c_{sm}=\sqrt{T_{em}(1+T_{i0}/T_{e0})/m_i}$. We note that these are only simple choices for a first attempt at this problem to limit the severity of initial transients. They are lacking in notably two ways: first, they lead to $\phi(t=0)=0$ via the field models described in sections~\ref{sec:linpol}-\ref{sec:adiabatic} while we know to expect a potential several $T_e/e$ at equilibrium, and second, the Maxwellian velocity dependence does not incorporate some basic kinetic expectations such as anisotropy and loss cones. Future simulations will use improved ICs as discussed in section~\ref{sec:compfuture}.

Lastly, to close the model in equations~\ref{eq:gkeq1d}-\ref{eq:sourceT} a field equation for $\phi$ is needed. Three different field models were employed to build confidence in the results and for computational reasons explained in section~\ref{sec:comp}. We also tried to use a fourth, zero-current field model akin to what was originally proposed for kinetic MHD~\cite{Kulsrud1983}, where the parallel electric field is obtained via a sum of the first moment of our kinetic equations~\ref{eq:gkeq1d} and assuming vanishing parallel currents. This model results in the field equation:
\begin{equation}
    -\sum_s\frac{q_s^2 n_s}{m_s\BparS}\pd{\phi}{z} = \sum_sq_s\left[ \pd{}{z}\left(\frac{M_{2\parallel s}}{\BparS}\right) + \frac{1}{2B}\pd{B}{z}\frac{M_{2\perp s}}{\BparS}\right], \\
\end{equation}
with the moments
\begin{eqnal}
    M_{2\parallel s} &= \int\dvw~\vpar^2\BparS f_s, \\
    M_{2\perp s} &= \int\dvw~(2\mu B/m_s)\BparS f_s.
\end{eqnal}
However, we did not find a stable numerical method for integrating the gyrokinetic model with this field equation.
For the moment, we have proceeded with the three field models described below because they worked and represent challenges that persist in 3D, which we wish to first understand and overcome in 1D.

\subsubsection{Kinetic e$^-$ with linear polarization} \label{sec:linpol}

In the spirit of taking a 1D limit of our long-wavelength gyrokinetic model, we create a proxy of the quasineutrality equation~\ref{eq:poisson} by replacing the nablas with a user-defined $i\kperp$ parameter. This is an approach that has been used in, for example, explorations of an ELM pulse in the scrape-off layer of tokamaks~\cite{Shi2015}. Later, we also show a scan of the assumed value of $\kperp$.
Furthermore, we use a linear polarization or Boussinesq approximation to replace the $n_s/B^2$ factor in the polarization density with the homogeneous, constant value $n_0/B_p^2$. Therefore in this field model, we compute the potential using
\begin{equation} \label{eq:poisson1dlinear}
    \kperp^2\sum_s\frac{n_0m_s}{B_p^2}\phi = \sum_s q_sn_s.
\end{equation}
The use of constant factors in the polarization density appears in other full-$f$ gyrokinetic codes~\cite{Michels2021}, and is often done to simplify, amongst other things, the linear solver required for this operation (i.e. the stiffness or left-side matrix becomes time-independent and is only factorized once). When this field model is used, we also drop the last term in equation~\ref{eq:hamiltonian}, as it is not needed to have a consistent, energy preserving scheme (see appendix~\ref{sec:energyConserv}).

\subsubsection{Kinetic e$^-$ with nonlinear polarization} \label{sec:nonlinpol}

When the constraint on the time step needed for time integration due to mirror forces is mitigated, the time step can then be set by the $\omega_H$ or electrostatic shear Alfv\'en mode (see section~\ref{sec:comp}). In order to ameliorate such constraint while continuing to obtain sufficiently high potential profiles, we tried to use a field model similar to that in equation~\ref{eq:poisson1dlinear} but with a larger $\kperp$ and keeping the spatial and temporal dependence of the density factor in the polarization density. That is, the potential was obtained from
\begin{equation} \label{eq:poisson1dnonlinear}
    \kperp^2\sum_s\frac{n_sm_s}{B_p^2}\phi = \sum_s q_sn_s.
\end{equation}
For the model to conserve energy, the $E\times B$ energy term in the Hamiltonian needs to be kept (equation~\ref{eq:hamiltonian}) and computed using $\abs{\v{v_E}}^2=(\kperp\phi/B_p)^2$ (see appendix~\ref{sec:energyConserv}). For higher accuracy, one could also keep the spatial dependence of $B(z)$ and $\kperp(z)$, using a proxy model $\kperp\sim\pi/a(z)\propto\sqrt{B(z)}$, where $a$ is the plasma minor radius. But the dominant variation in the polarization term is the density's, as we will see in section~\ref{sec:results}, so we used $\kperp/B^2=\mathrm{constant}$ for now.

\subsubsection{Boltzmann, isothermal e$^-$ with ambipolar sheath} \label{sec:adiabatic}

This model follows earlier work~\cite{Shi2015}, which assumes that electrons have Boltzmann behavior and are isothermal, such that when we impose ambipolarity at the sheath entrance (equality of electron and ion particle fluxes), one obtains an equation for the sheath potential
\begin{equation} \label{eq:phi_s}
    \phi_s = \phi(z=\pm\Lz/2) = \frac{T_{e0}}{e}\ln\left(\frac{n_iv_{te0}}{\sqrt{2\pi}\,\Gamma_i}\right)\Bigg|_{z=\pm\Lz/2}
\end{equation}
in terms of the kinetic ion fluxes ($\Gamma_i$) at the sheath entrances $z=\pm\Lz/2$ for the computational domain $z\in\left[-\Lz/2,\Lz/2\right]$. Equation~\ref{eq:phi_s} assumed quasineutrality and used $v_{te0}^2=T_{e0}/m_e$. Having obtained the sheath potential, quasineutrality and a Boltzmann electron response, $n_e=C\exp\left[e(\phi-\phi_s)/T_{e0}\right]$ (where $C$ is a constant determined by $n_i(z=\pm\Lz/2)=n_e(z=\pm\Lz/2)$), combine assuming zero wall potentials to produce the field equation
\begin{equation} \label{eq:phiadiabatic}
    \phi = \phi_s + \frac{T_{e0}}{e}\ln\left(\frac{n_i}{n_{is}}\right),
\end{equation}
where $n_{is}$ is the ion density at the sheath entrance. Only the ion gyrokinetic equation needs to be evolved in this model, roughly increasing the time step needed in time integration by roughly $\sqrt{m_i/m_e}\sim60$ and making the calculations much cheaper.

Note that electrons trapped inside the mirror bounce back and forth many times and are confined long enough to thermalize, developing a Maxwellian velocity dependence and exhibiting a Boltzmann response to the potential. But in the expander $f_e$ has a significant beam-like population of cold electrons that escaped the mirror. This population gives $f_e$ a non-Maxwellian character and may reach the wall before they have time to thermalize; i.e. the expanders are collisionless and not expected to follow a Maxwellian distribution~\cite{Ryutov2005}. The validity of this Boltzmann electron model is thus weak in \wham's expanders, and we merely employ it as a tool to make iterative exploration possible (given its lower cost) and to help us understand the computational challenges of modeling ion dynamics in high-field mirrors. 

\section{Computational challenges and techniques} \label{sec:comp}

The gyrokinetic model in equations~\ref{eq:gkeq1d}-\ref{eq:nTicsL} with the field models in sections~\ref{sec:linpol}-\ref{sec:adiabatic} are discretized with a discontinuous Galerkin (DG) method and time-integrated with an explicit strong-stability-preserving third order Runge-Kutta scheme in the \gkeyll~code~\cite{gkeyllWeb}. The spatial discretization is carried out on a multidimensional, uniform grid over the domain $(z,\vpar,\mu)\in\left[-\Lz/2,\Lz/2\right]\times\left[-m_s\vparmax^2/2,m_s\vparmax^2/2\right]\times\left[0,\mumax\right]$ using $\Nz\times\Nvpar\times\Nmu$ cells with a piecewise polynomial basis of order $p$ within each cell. We will primarily discuss results obtained with $(\Nz,\Nvpar,\Nmu)=(288,64,192)$ and $p=1$, corresponding to ${>}28$ million degrees of freedom per species, $z$ limits obtained from equation~\ref{eq:zofZ} for $Z\in[-2.5\,\mathrm{m},2.5\,\mathrm{m}]$, $\vparmax=3.75v_{ts0}=3.5\sqrt{T_{s0}/m_s}$ and $\mumax=m_s(3v_{ts0})^2/(2\Bp^2)$. These velocity grids are marginally sufficient, as discussed below.

A couple of features motivate high velocity-space resolution. One reason seemingly extreme $\mu$ resolution is the location of the trapped-passing boundary. Consider that particles at this boundary in the center of the mirror ($z=0$) with $\vpar=v_{ts0}$ have $\mu=\mu_{tp}=m_sv_{ts0}^2/(2\Bm)$ in the large $\Rm$ limit. Defining a thermal $\mu=\mu_t=m_sv_{ts0}^2/(2\Bp)$ at $z=0$, we see that the trapped-passing boundary $\mu_{tp}$ is quite small by comparison: $\mu_{tp}/\mu_t=\Bp/\Bm=1/20$ for the Pleiades \wham~case in figure~\ref{fig:eqcomp}. For the choice of $\mu_{\mathrm{max}}=9\mu_t$ to handle the high velocity part of a Maxwellian, and a grid spacing of $\Dmu=\mu_{tp}$ to resolve the trapped passing boundary, we see that $\Nmu=180$ cells or more are needed.
While this $\Dmu=\mu_{tp}$ is marginal, this is alleviated some by the fact that a DG algorithm with $p$ order polynomials is equivalent to $p+1$ optimally-located Gaussian quadrature points per cell in each direction. So the resolution capability of a $p=1$ DG code with $N$ cells is comparable to a low-order grid with approximately $2N$ grid points.
Also, as the electrostatic potential builds up to confine electrons, the trapped passing boundary shifts to higher energy, which is easier to resolve.

The combination of uniform velocity grids and large temperature variation is another feature calling for high velocity resolution. Typically continuum simulations have fixed (in $\v{x}$) velocity grids with extents $\vparmax\geq4v_{ts0}$ (and an equivalent $\mumax$). For a plasma with temperatures that vary weakly in $z$, such grids are sufficient to resolve $f_s$ well at all $z$ and maintain $f_s$ small near the boundary by a factor of $\lesssim\exp\left[-\vparmax^2/(2v_{ts0}^2)\right]\sim10^{-4}$, so we can safely employ zero-flux BCs there. An HTS mirror, unlike the tokamak core that gyrokinetic models and codes have historically been developed for, has large temperature gradients along the field line. This means that a velocity grid that may seem sufficient in the cool expander would be far too small in extent for the plasma at $z=0$ or vice-versa, a velocity grid that is appropriate in the center would be far too large and coarse for the expander. Continuum gyrokinetic codes, like \gkeyll, are then forced to either use a large grid with many cells at all $z$, or to compromise with grids that are smaller and more over-resolved than necessary at $z=0$, and are fine enough to provide a reasonable description near and past the throat without being overly expensive. This compromise resulted in the grids used in this work.

It is worth interjecting with the caveat that two other approaches to support large temperature variations in kinetic simulations have been pursued elsewhere. One of them is to use multiblock grids where each block along $z$ has a different $\vparmax$ or resolution~\cite{Jarema2016}. The second technique involves mapping velocity coordinates to ones normalized by the local thermal speed, e.g. $\widehat{v_{\parallel}}=\vpar/v_{ts} (z)$ and $\hat{\mu}=\mu/(m_sv_{ts}^2(z)/(2B_p))$~\cite{Taitano2016}. Although these strategies may be further developed to work for our purposes, they can both suffer from diffusion in the $\mu$ direction either due to the inter-block interpolations or the numerical diffusion in the fluxes along $\mu$ that arise from the extra terms that normalizing $\mu$ introduces. Such diffusion would violate exact $\mu$ conservation, which is paramount in accurately depicting mirror processes. Additionally, multiblock velocity grids have only been used in one $\delta f$ code, leaving us with the uncertainty of how boundaries in the direction of other blocks but not abutting other blocks must be treated in full-$f$ codes.

The compromise that we have made for the grids is only half of the challenge; the other half is the duration that one intends to simulate and what time step ($\Dt$) the numerical methods call for (thus determining the number of time steps needed). We leave the latter question for section~\ref{sec:dtconstraints} below and discuss two time scales of interest. As stated in section~\ref{sec:intro}, one of our primary objectives is to determine whether our gyrokinetic model is consistent with a Pastukhov equilibrium. That is because in the future, when we study microstability and perpendicular transport, we would like to start from such an equilibrium and evolve the turbulence without potentially problematic macro-rearrangements of the plasma caused by the search for a different equilibrium. The Pastukhov potential, however, arises from a difference in the electron and ion collision times, which for the parameters in this work are nominally $\tau_{ee}\simeq 18\,\mu\mathrm{s}$ and $\tau_{ii}\simeq 36\,\mathrm{ms}$. Thus, to form this potential from a state away from equilibrium could take at least several $\tau_{ee}$ or a $\tau_{ii}$, which for the small $\Dt$ required could take millions to billions of time steps. In the following subsections, we describe the origin of this small $\Dt$, and some ways to overcome it, yet reiterate that in the future, we will begin from a Pastukhov equilibrium estimated elsewhere (e.g. with semi-analytic methods or bounce-averaged Fokker-Planck codes).

\subsection{Time step constraints} \label{sec:dtconstraints}

Given that the collisionality is relatively low for the plasmas of interest, the two primary constraints on $\Dt$ arise from the need to resolve the mirror acceleration and the electrostatic shear Alfv\'en (or $\omega_H$) mode, the latter only arising in kinetic electron simulations. When integrated with explicit methods, the mirror acceleration, contained in the third term on the left side of equation~\ref{eq:gkeq1d}, imposes the Courant-Friedrichs-Lewy (CFL) stability limit 
\begin{equation} \label{eq:mirrorCFL}
    \frac{\Dt}{\Dvpar/(2p+1)}\max\left[\frac{1}{m_s}\pd{(\mu B)}{z}\right] \lesssim C,
\end{equation}
where $\Dvpar=2\vparmax/\Nvpar$ is the cell length in $\vpar$ and $C$ is an order unity number. Thus, the fine $\vpar$ grids needed to resolve $f_s$ near and beyond the mirror throat, the small electron mass, the large $\mumax$ needed to capture $f_s$ in the center, and the large gradients in $B$ due to the HTS coils all conspire to give a small $\Dt$. For the physical and computational parameters in sections~\ref{sec:models}-\ref{sec:comp}, equation~\ref{eq:mirrorCFL} yields 
$\nu_{ee}\Dt\simeq 1/(6\times10^6)$, meaning that around 6 million time steps with large grids would be needed to reach a single $\tau_{ee}$, and many more to build the equilibrium potential from $\phi=0$.

When the limitation in equation~\ref{eq:mirrorCFL} is overcome, say by using an implicit method or with the technique described in section~\ref{sec:forcesoft}, Boltzmann electron simulations are primarily limited by the parallel ion advection CFL, i.e.
\begin{equation}
    \max\left(v_{\parallel i}\kpar\right)\Dt \lesssim C,
\end{equation}
typically an easily met criteria ($\kpar$ is the wavenumber along the field line). On the other hand, kinetic electron simulations are limited by the need to resolve the $\omega_H$ mode, the electrostatic branch of the shear Alfv\'en wave~\cite{Lee1987}. In the uniform slab with a linearized polarization, this constraint reads
\begin{equation} \label{eq:omegaHlinCFL}
    \omega_H\Dt = \frac{v_{te0}k_{\parallel\max}}{\kperp\rho_{s0}}\sqrt{\frac{n_{e0}(z,t)}{n_0}}\Dt \lesssim C.
\end{equation}
Here $\rho_{s0}=c_{se0}/\omega_{ci}$ is the reference ion sound gyroradius, with $c_{se0}=\sqrt{T_{e0}/m_i}$ and $\omega_{ci}=q_i\Bp/m_i$ the ion gyrofrequency. The linear analysis that leads to the frequency in equation~\ref{eq:omegaHlinCFL} is done around the instantaneous and local equilibrium density $n_{e0}(z,t)$ (see appendix~\ref{sec:omegaHlin}). This high frequency also becomes harder to resolve as we refine the $z$-grid, which is needed in order to capture the field-aligned variation near and outside of the mirror throats since $k_{\parallel\max}\simeq \pi(p+1)/\Dz = \pi(p+1)\Nz/\Lz$.

As we will see in section~\ref{sec:results}, there is some indication that decreasing $\kperp$ can produce larger potential drops that are closer to the Pastukhov level and the adiabatic electron result. However, a decrease in $\kperp$ entails a drop in $\Dt$, if the mirror-force constraint is mitigated, and can make the simulation intolerably expensive. If $\kperp\rho_{s0}=0.01$, then our base parameters produce an $\omega_H$ frequency with a time step requirement, $\nu_{ee}\Dt\simeq 1/(8.3\times10^6)$, which is quite close to that demanded by the mirror force.


We are considering semi-implicit methods to bypass equation~\ref{eq:omegaHlinCFL}, but in the meantime, we have instead opted to utilize a larger $\kperp$ and instead obtain a suitably peaked potential by incorporating the variation of the density in the  polarization as described in section~\ref{sec:nonlinpol}. In that case, the $\omega_H$ constraint becomes (see appendix~\ref{sec:omegaHnonlin})
\begin{equation} \label{eq:omegaHnonlinCFL}
    \omega_H\Dt = \frac{v_{te0}\kpar}{\kperp\rho_{s0}}\sqrt{\frac{n_{e0}(z,t)}{n_{i0}(z,t)}}\Dt  \lesssim C.
\end{equation}
This $\sqrt{n_{e0}(z,t)/n_{i0}(z,t)}$ factor is everywhere very close to 1, in contrast to equation~\ref{eq:omegaHlinCFL} where the $\sqrt{n_{e0}(z,t)/n_0}$ factor drops off in the expander. At first glance, this change would indicate that the nonlinear polarization model would need a smaller $\Dt$, and all things being equal, it does. However, in the confined plasma where the pressure is largest, these factors are both close to 1 and the $\sqrt{n_{e0}(z,t)/n_{i0}(z,t)}$ factor in equation~\ref{eq:omegaHnonlinCFL} will be offset by our increasing $\kperp$ tenfold. We intend for this larger $\kperp$ to produce reasonable $\phi$ profiles still, thanks to the natural decrease in $n_s(z,t)$ away from the center introduced into the polarization equation~\ref{eq:poisson1dnonlinear}. These changes ought to lead to the friendlier approximate time step $\nu_{ee}\Dt\simeq 1/(8.3\times10^5)$.

\subsection{Mitigating the mirror $\Dt$ by force softening} \label{sec:forcesoft}

\begin{figure}
\includegraphics[width=0.48\textwidth]{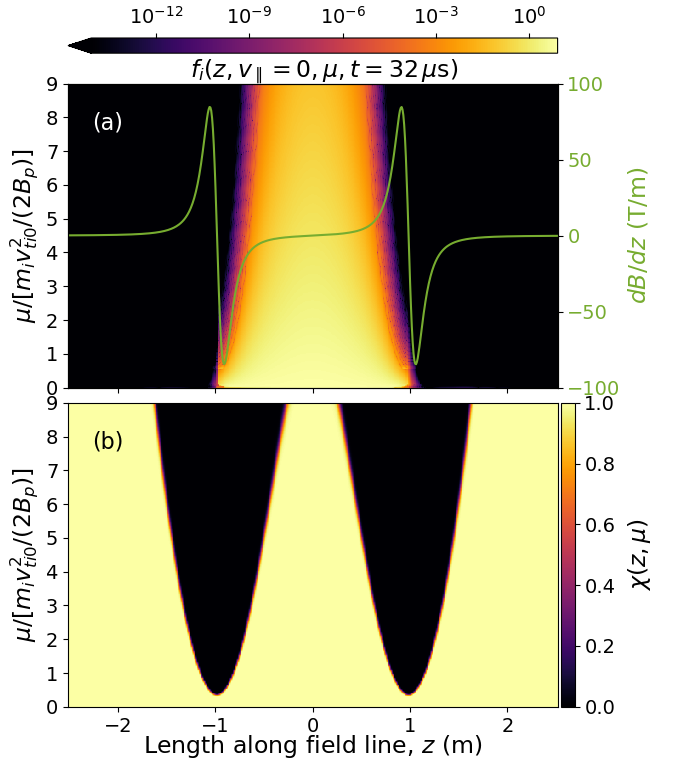}
\caption{\label{fig:fiadiab128log} (a) Ion distribution function at $\vpar=0$ and $t=32\,\mu\mathrm{s}$ (negative cells multiplied by -1 for simplicity of viewing) and the magnetic field gradient (green line). (b) Softening function $\chi(z,\mu)$ in equation~\ref{eq:softChi}.}
\end{figure}

Upon closer examination of equation~\ref{eq:mirrorCFL} we notice that the regions of phase space that call for the smallest $\Dt$ are those electron ones (due to their small $m_s$) with large $\mu$ and $\partial_z B$. The magnetic field gradient is maximized as one approaches the mirror throat, but at the same time, particles with larger $\mu$ are reflected by mirror forces. So in those same regions where $\max[\partial_z(\mu B)]$ occurs, we expect $f_s$ to be very small (figure~\ref{fig:fiadiab128log}(a)).
It is thus unfortunate that explicit time integration demands a tiny $\Dt$ from a region in which $f_s$ is many orders of magnitude smaller and makes a possibly negligible contribution to the system's dynamics.

One way to mitigate this CFL constraint is to soften the mirror force in those regions of negligible $f_s$, inspired by the way high-speed particles are slowed down in some particle-in-cell methods~\cite{Werner2018}. Our approach here, to maintain the Hamiltonian structure of the system, is to introduce a softened Hamiltonian
\begin{equation} \label{eq:softHamiltonian}
    H = \frac{1}{2}m_s\vpar^2 + \Pi(z,\mu)~\mu B + q_s\phi - \frac{1}{2}m_s\abs{\v{v_E}}^2,
\end{equation}
where we multiplied the term giving rise to the mirror force by a phase-space factor $\Pi(z,\mu)$. We intend to use this factor to modify the advection speed in the $\vpar$ direction so that the mirror force term, given the Poisson bracket in equation~\ref{eq:poissonBracket1d},
\begin{eqnal}\label{eq:softVparDot}
    \dot{\vpar}_{\mathrm{mirror}} = -\frac{1}{m_s\GJac\BparS}\pd{}{z}\left(\Pi(z,\mu)\mu B\right) = -\frac{\chi(z,\mu)}{m_s\GJac\BparS}\mu\pd{B}{z}
\end{eqnal}
is weakened by a factor $\chi(z,\mu)$ of our choosing. Once $\chi(z,\mu)$ is defined, the Hamiltonian softener $\Pi(z,\mu)$ can be obtained by integrating equation~\ref{eq:softVparDot} from $z=0$ with the boundary values $\Pi(z=0,\mu)=\chi(z=0,\mu)=1$. A weaker $\mu B$ term in the Hamiltonian changes the CFL constraint due to the mirror forces from equation~\ref{eq:mirrorCFL} to
\begin{equation} \label{eq:softMirrorCFL}
    \frac{\Dt}{\Dvpar/(2p+1)}\max\left[\frac{\chi(z,\mu)}{m_s}\pd{(\mu B)}{z}\right] \lesssim C.
\end{equation}

The $\chi(z,\mu)$ factor is chosen to remain $1$ where $f_s$ is appreciable but vanishes in regions limiting $\Dt$ despite their negligible $f_s$. One possibility is to choose $\chi(z,\mu)=1$ in most places of interest and $\chi(z,\mu)\propto \mu_t(z)/\mu$ where $f_s$ is negligibly small, with $\mu_r(z)$ being a threshold adiabatic moment beyond which we soften the Hamiltonian.   However, $\chi(z,\mu)$ decays too slowly in that case, so we instead opt for
\begin{eqnal} \label{eq:softChi}
    \chi(z,\mu) = \begin{cases}
        1 \qquad &\mu\leq\mu_t(z), \\
        \frac{1}{2}\left[1+\tanh\left(\frac{\mu_t(z)+2.5w-\mu}{w/2}\right)\right] \qquad &\mu>\mu_t(z),
    \end{cases}
\end{eqnal}
with $w=\mu_t(z)/10$. We define $\mu_t(z)$ as the largest adiabatic moment that is accessible at each $z$, $\mu_t(z) = \mu_{max} B_p / B(z)$, according to energy conservation but disregarding collisions, parallel kinetic energy (a relatively small correction for this high-field mirror) and the electrostatic potential because we do not know it a priori. 



The softening function $\chi(z,\mu)$ in equation~\ref{eq:softChi}, used for some results in section~\ref{sec:results}, is plotted in figure~\ref{fig:fiadiab128log}(b). Notice, by comparison with figure~\ref{fig:fiadiab128log}(a), that the regions where the distribution function is many orders of magnitude smaller coincide with regions where $\chi(z,\mu)\to0$. This correspondence is not exact, since other physics, like electrostatic forces, also shape the regions of small $f_s$. Additionally, electrostatic acceleration is signed and modifies the regions of small $f_i$ and $f_e$ differently. Therefore the optimal $\chi(z,\mu)$ will look slightly different for each species. In the future, we can incorporate estimates of the potential profile $\phi(z)$ in the definition of $\mu_t(z)$ to produce a more suitable softening function. Nonetheless, in section~\ref{sec:results}, we show results obtained with this $\chi(z,\mu)$ are sufficiently close to those without force softening and were obtained 19.5 times faster.

\subsection{Further computational improvements} \label{sec:compfuture}

For Boltzmann electron simulations, force softening is sufficient to attain large time steps that result in simulations where the equilibrium (Pastukhov) potential is built from scratch (i.e. $\phi(t=0)=0$) in a matter of hours (on 288 cores). This is a nearly 20X speedup over simulations without force softening. However, kinetic electron simulations can still have a $\Dt$ small enough to be impractical for parameter scans. Therefore, additional measures must be taken to further reduce the cost of these simulations if kinetic electron physics is of interest.

Assuming that it is, there are five modifications to the approach presented in this work that we are currently pursuing. These are:
\begin{enumerate}[leftmargin=12pt]
    \item Simulation time: As mentioned earlier and illustrated in section~\ref{sec:results}, developing the (Pastukhov) equilibrium potential from $\phi(t=0)=0$ can take from several $\nu_{ee}^{-1}$ to $\sim\nu_{ii}^{-1}$, which could require millions or billions of time steps with large grids and costly simulations. However, the ultimate goal of mirror gyrokinetics is to examine the stability of gradient-driven modes (e.g. interchange) and perpendicular turbulent transport, which occurs on a much smaller time scale than that of the equilibrium evolution. For example, for the parameters used in this work, the potential takes on the order of $100\,\mu\mathrm{s}$ to reach the analytic estimate, while the interchange time scale is
    $\sim\zm/(2v_{ti0})\lesssim 1\,\mu\mathrm{s}$~\cite{Fowler1981}.
    Therefore, studies of interchange stability and transport would time integrate for a $\sim$10X smaller period and result in $\sim$10X cheaper simulations (assuming perfect perpendicular parallelism).
    \item Improved ICs: For both 1D and higher-dimensional simulations, one could start from ICs closer to the final (Pastukhov) equilibrium than those in equations~\ref{eq:nTics}-\ref{eq:nTicsL} to shorten the required simulation time. 
    Such ICs could be obtained with semi-analytical~\cite{Egedal2022} or analytical (under development) methods that compute the equilibrium more quickly. Another approach is to first perform a Boltzmann electron simulation, which is $\sim\sqrt{m_i/m_e}\approx 60$ times faster, and use that result as the IC of a kinetic electron simulation. 
    \item Nonuniform grids: the cost of these simulations can be reduced by using nonuniform grids with higher resolution only where needed. For example, in \gkeyll~one could use a coordinate mapping $(z,\vpar,\mu)=(z(\xi),\vpar,\mu(\eta))$, where $\xi$ and $\eta$ are the computational coordinates along the field line and along $\mu$, respectively, to place more degrees of freedom (DOFs) near the mirror throats and at low $\mu$ (DOFs uniformly spaced in $\xi$ and $\eta$). 
    We estimate that this could reduce the number of DOFs, and thus speed up the code, by a factor of 10-15 -- a factor of 2-3 from using a nonuniform $z$ grid and a factor of 5 from a proportional $\mu$ grid
    \footnote{For example, consider a $\mu$ grid that has the same grid spacing as a uniform grid at small $\mu$ but transitions to a relative spacing grid at higher $\mu$, of the form $\Delta \mu_j = {\rm max}(r \mu_{j-1/2}, \Delta \mu_0)$. This relative grid with a 12\% relative grid spacing ($r=0.12$) and just 36 cells can span the same range as a uniform grid with 192 cells.}.
    \item Positivity preservation: A novel, rigorous method for preserving positivity of the particle distribution function while conserving energy in Hamiltonian systems has been developed~\cite{Mandell2021thesis} and implemented in another branch of \gkeyll. Ensuring $f_s\geq0$ would improve robustness of simulations on coarser grids, and thus reduce computational cost.
    \item Improved parallelization: The results in this manuscript were produced with an earlier version of \gkeyll; since then, additional parallelization strategies have been implemented. Specifically, these simulations can now be performed on GPUs, parallelizing over velocity space and species. Parallelization over species alone reduces the time to solution by 2X, and similar reductions are found from GPUs and velocity space parallelization. 
\end{enumerate}

Most of these improvements are ready for use. Hence we expect that future work will take advantage of such capabilities to perform simulations at a small fraction of the cost of the simulations reported here. Additionally, we may use such enhancements to do more complex studies like parameter scans, 3D simulations, or simulations with sloshing ions.

\section{Results} \label{sec:results} 

In this section, we report on results from three types of simulations we carried out; namely those using
\begin{itemize}
    \item Boltzmann electrons with \& without force softening,
    \item kinetic electrons with linearized polarization without force softening, and
    \item kinetic electrons with nonlinear polarization \& force softening.
\end{itemize}
These results can be reproduced using the \gkeyll~input files found under version control at \url{https://github.com/ammarhakim/gkyl-paper-inp/tree/master/2023_PoP_gkwham1x2v}. As stated earlier, our two-fold intention here is to examine the computational challenges in modeling an HTS mirror with a continuum gyrokinetic code and to determine if these models develop the (Pastukhov) electrostatic potential expected for such machines.

We can obtain an approximation for the potential drop between the center of the plasma and the mirror throat by considering the collisional electron loss rate computed by Pastukhov~\cite{Pastukhov1974} and later corrected by others~\cite{Chernin1978,Cohen1978}:
\begin{equation} \label{eq:dnedt}
    \d{n_e}{t} = - \frac{4}{\sqrt{\pi}}A_p\frac{\nu_{ee}n_e}{G(\Rm)}\left(\frac{e\phi}{T_{e0}}\right)^{-1}e^{-e\phi/T_{e0}}\,I(T_{e0}/e\phi)
\end{equation}
in terms of the functions
\begin{eqnal}
    G(x) &= \frac{2x+1}{2x}\ln(4x+2), \\
    I(x) &= 1+0.5\sqrt{\pi\,x}\,e^{1/x}\mathrm{erfc}\left(\sqrt{1/x}\right).
\end{eqnal}
We have introduced a scaling factor $A_p$ for several reasons. First, we want to observe the change in the potential drop as one scans the collision frequency. Second, the Pastukhov-Chernin-Cohen approximation incorporates electron-electron and electron-ion collisions, but our kinetic electron simulations only account for electron-electron collisions for now. 
By itself, this would make it more appropriate to compare our kinetic results with the analytic Pastukhov estimate using a scaling factor of $A_p=0.5$, but there are other effects that may make the effective $A_p$ greater than 1.
There is also an issue with differences between the velocity dependence of the collision frequency in the Dougherty (this work) and LRO (used by Pastukhov) operators, as explained in section~\ref{sec:gkmodels}. These effects would increase $A_p$, though the final impact on the potential depends on the relative importance of pitch-angle and velocity scattering for electrons and ions.  Future code validation will employ the equivalent of Pastukhov's calculation using the Dougherty operator (under development), but for now we see below that the exact results depend weakly on $A_p$.

\begin{figure}
\includegraphics[width=0.48\textwidth]{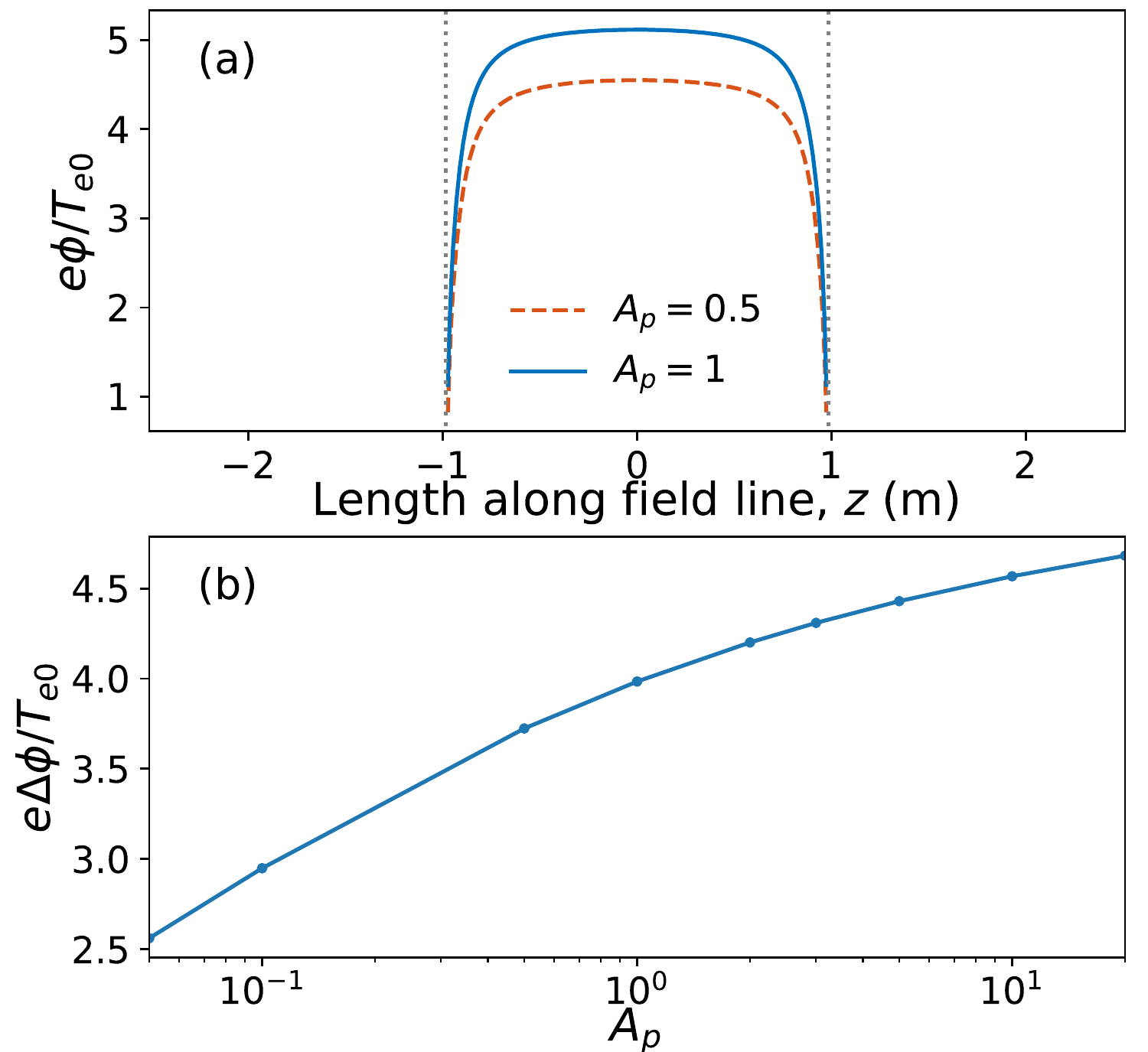}
\caption{\label{fig:pastukhovPhi}(a) Estimated Pastukhov potential profile for the mirror using parameters in tables~\ref{tab:eqpars}-\ref{tab:simPars} and equations~\ref{eq:dnedt}-\ref{eq:taupei}, absent in the expanders because the assumptions are not valid there. Vertical dotted lines indicate the mirror throat locations. (b) Potential drop between the mirror center ($z=0$) and the mirror throat ($z=z_m=0.98~\mathrm{m}$) in the Pastukhov estimate as a function of an electron collision frequency scaling factor ($A$): $\nu_{ee}\to A\,\nu_{ee}$.}
\end{figure}

From equation~\ref{eq:dnedt} we may calculate an electron particle confinement time $\tau_{pe}=n_e/(dn_e/dt)$. The ion particle confinement time, however, is much longer because ions are collisionally scattered less frequently and is of the order $\tau_{pi}=\nu_{ii}^{-1}\log\Rm$ according to analytical and numerical study of the LRO equation~\cite{Baldwin1977}. Since electrons are scattered into the loss cone and thus exit to the expander and the sheath more quickly, they leave behind the ions, and an electrostatic potential begins to build. This potential starts to decelerate electrons and accelerate ions until the loss rates for both species equalize and
\begin{equation} \label{eq:taupei}
    \tau_{pe} = \tau_{pi}.
\end{equation}
Said differently, a potential develops to maintain quasineutrality. The so-called Pastukhov potential $\phi$ the satisfies equation~\ref{eq:taupei} may be numerically solved for via root finding. For the parameters in tables~\ref{tab:eqpars}-\ref{tab:simPars}, we obtain the potential given in figure~\ref{fig:pastukhovPhi}(a). The assumptions that lead to equation~\ref{eq:dnedt} are not valid in the expander due to, for example, the large mean flow velocities and weak collisionality therein; for this reason figure~\ref{fig:pastukhovPhi}(a) only shows $\phi$ inside the mirror.

A key metric for mirrors is the potential drop between the central plasma ($z=0$) and the mirror throat ($z=\zm=0.98\mathrm{m}$), since a sufficiently large electric field is needed to elecrostatically confine the electrons and reduce parallel electron heat convection. In this Pastukhov estimate, the potential drop normalized to reference electron temperature is $e\Delta\phi/T_{e0}=3.98$ (3.72) for $A_p=1$ (0.5), which is in the range of those numerical and experimental results mentioned in section~\ref{sec:intro}. Additionally, since this potential arises from the difference in scattering rate between ions and electrons, if the electrons scatter more frequently (increasing $\nu_{ee}$), a larger potential drop will develop to contain them and maintain quasineutrality. This phenomenon explains the trend in figure~\ref{fig:pastukhovPhi}(b), for which we computed the potential drop corresponding to a series of scaled electron collision frequencies. The increase in $e\Delta\phi/T_{e0}$ with $\nu_{ee}$ will partially explain some differences between Boltzmann and kinetic electron simulations presented below.

\subsection{Boltzmann electrons} \label{sec:resultsBoltzmannElc}

\begin{figure}[b]
\includegraphics[width=0.48\textwidth]{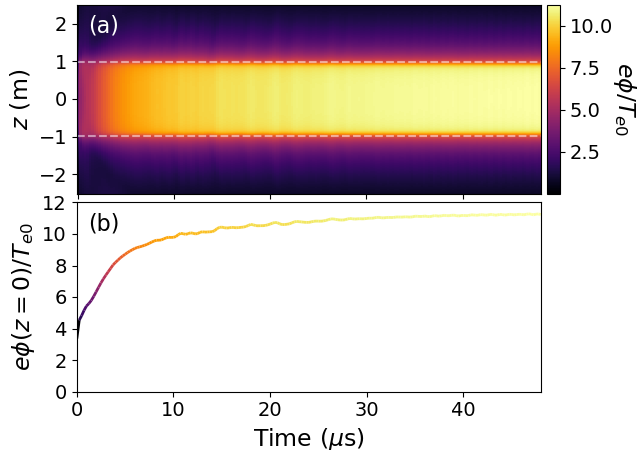}
\caption{\label{fig:adiabaticPhi}(a) Electrostatic potential $\phi(z,t)$ in the case with Boltzmann electrons (mirror throats indicated with white dashed lines), and (b) central electrostatic potential for $t=0-48\,\mu\mathrm{s}=0-(8/3)\nu_{ee}^{-1}$.}
\end{figure}

The Boltzmann electron model evolves the ion distribution function $f_i(z,\vpar,\mu)$ according to equation~\ref{eq:gkeq1d} and the field model in section~\ref{sec:adiabatic}. The initial condition consists of a Maxwellian $f_i$ and has $\phi(t=0)=0$ because no ion fluxes have crossed the sheath entrance at that time yet. As we integrate the model in time, ions are lost from the central region to the expander and sheath via pressure relaxations and collisional scattering. The ions in the expander are also further accelerated by the mirror force pointing towards the sheath, where they are lost by the sheath BCs that absorb all arriving ions. The result is an ion density that decreases away from the central region and gives rise to a positive potential profile via equation~\ref{eq:phiadiabatic}. The potential that develops works against the mirror force to accelerate ions through the mirror throat and towards the sheath, thus enhancing the loss of ions. These losses are eventually balanced by the plasma source near $z=0$, at which point the simulation reaches a steady state.

\begin{figure}[b]
\includegraphics[width=0.48\textwidth]{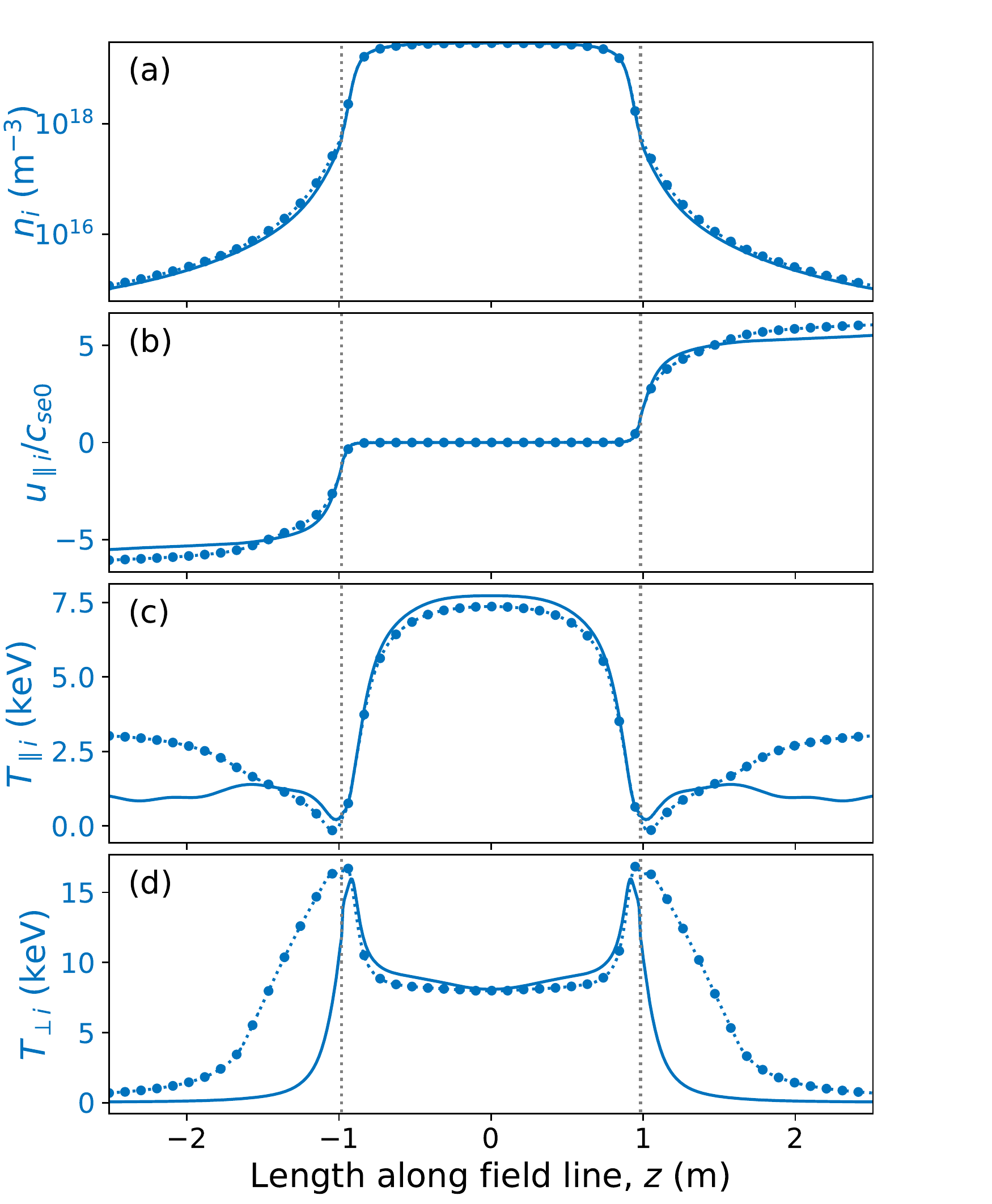}
\caption{\label{fig:adiabaticMoms}Ion guiding center density ($n_i$), mean drift speed ($\upari$) normalized to $c_{se0}=\sqrt{T_{e0}/m_i}$, and parallel and perpendicular temperatures ($\Tpari$, $\Tperpi$) at $t=32\,\mu\mathrm{s}\approx1.78\nu_{ee}^{-1}$ in Boltzmann electron simulation without (solid) and with (dotted with circles) force softening. Vertical dotted lines indicate the mirror throat locations.}
\end{figure}

Reaching this steady state of the ion profiles likely requires time-integrating the system to $O(\nu_{ii}^{-1})$ seconds. Our goal, however, is only to integrate it long enough to build the bulk of the electrostatic potential. We did so for nearly $4\nu_{ee}^{-1}$ at first, requiring 73.52 hours on 288 cores when using the $288\times64\times192$ grids mentioned in section~\ref{sec:comp} that were chosen as a result of a resolution scan described in appendix~\ref{sec:resconv}. The spatio-temporal evolution of $\phi$, for example, is shown in figure~\ref{fig:adiabaticPhi}. We see that a peaked potential is formed almost immediately and continues to rise in amplitude as one approaches $t=\nu_{ee}^{-1}=18\,\mu\mathrm{s}$. By the time $t\approx(8/3)\nu_{ee}^{-1}=48\,\mu\mathrm{s}$ is reached, the central potential appears to saturate around $e\phi(z=0)/T_{e0}=11.28$, hinting that we may have run the simulation long enough.

Near this possible steady state, at $t=32\,\mu\mathrm{s}\approx1.78\nu_{ee}^{-1}$, the ion fluid moments have developed to those in figure~\ref{fig:adiabaticMoms} (solid lines). The ion density remains peaked near the reference density $n_0$, demonstrating confinement of the plasma; the density drops by over 4 orders of magnitude from the center to the sheath entrance, a decrease much faster than $B(z)/\Bp$. The ion mean parallel flow speed in figure~\ref{fig:adiabaticMoms} exhibits a relative stagnation in the confinement region and a rapid acceleration to $\upari=c_{se0}$ near the mirror throats. This increase in $\upari$ continues in the expander due to combined contributions from the parallel electric field, the mirror forces and conservation of mass (e.g. as in a nozzle). In terms of the local sound speed, $c_s=\sqrt{(T_{e0}+3\Tpari)/m_i}$, the ions also become supersonic at $z=\zm$ but are instead accelerated to $\upari/c_s\approx2-3$ by the time they reach the sheath. This is reminiscent of the accelerating solution obtained for low (source) injection velocities in recent PIC simulations~\cite{Jimenez2022}. Lastly, figure~\ref{fig:adiabaticMoms} also demonstrates the strong anisotropy of this scenario throughout the whole device, especially near the throat where $\Tperpi$ is considerably larger than $\Tpari$ (40 times larger in fact) as one would expect from conservation of energy and $\mu$.

\begin{figure}[b]
\includegraphics[width=0.48\textwidth]{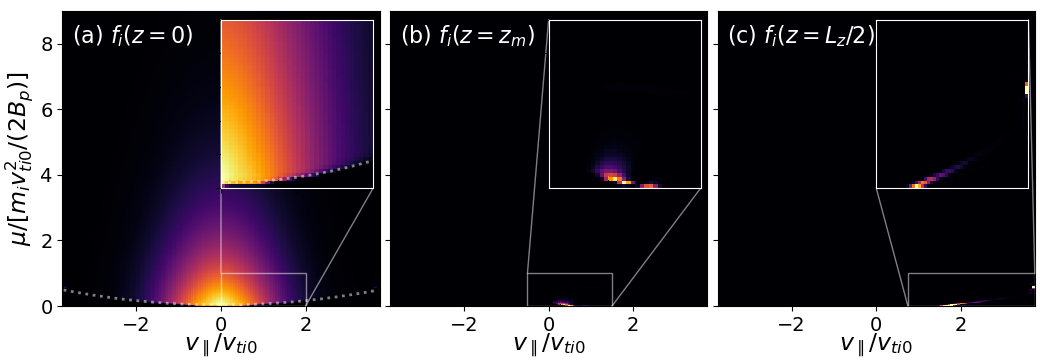}
\caption{\label{fig:adiabaticfi}Ion distribution function in Boltzmann electron simulation at $t=48\,\mu\mathrm{s}=8\nu_{ee}^{-1}/3$ at three locations along the field line: (a) the center, (b) the mirror throat, (c) the sheath entrance in the expander.}
\end{figure}

The temperature can be quite sensitive to the velocity-space details of the distribution function, and upon examination, we see why this Boltzmann electron simulation was not continued past $\nu_{ee}t=8/3$. Figure~\ref{fig:adiabaticfi} presents the ion distribution function at $\zm=\{0,\zm,\Lz/2\}$, i.e. the center, throat and sheath entrance. The first panel shows the archetypal mirror plot, except that due to the strong magnetic field in this device and the fact that it is shown in $\vpar-\mu$ coordinates, the loss cone appears narrow and parabolic. Nonetheless, the simulation's $f_i(z=0,\vpar,\mu)$ follows the (dotted white) contour of the trapped boundary given by
\begin{equation} \label{eq:trappedpassing}
    \frac{\mu\Bp}{\mu\Bp + m_s\vpar^2/2 + q_s\delta\phi} \geq \frac{1}{R_m},
\end{equation}
including the contribution of the self-consistently generated potential ($\delta\phi=\phi(z=0)-\phi(z=\zm)$) that leads to the ambipolar hole near the origin. As one approaches and passes through the mirror throat, only low-$\mu$ particles that were not mirror confined can be found, and $f_i$ becomes very localized near $\mu=0$. This makes it challenging to resolve $f_i$ in these regions if we use uniform grids 
, as \gkeyll~and some other gyrokinetic codes do. In this case, the distribution function is perhaps only marginally resolved at the throat, where $f_i$ is only appreciable in a few cells along $\mu$. In the last panel, located at the sheath entrance, we see an $f_i$ that slants towards higher $\vpar$ and $\mu$. This is a result of an acceleration of particles due to the Yushmanov potential (i.e. combined mirror and electrostatic potential). We can estimate the $\vpar$ to which particle just inside the loss cone at $(z,\vpar,\mu)=(0,2.5v_{ti0},0.2374m_iv_{ti0}^2/(2\Bp))$, where $f_i$ is only 100 times smaller than near the origin, is accelerated to by the time it reaches the sheath entrance from conservation of energy:
\begin{eqnal}
    \vpar &= 2.5v_{ti0}+\left[2\frac{e\delta\phi+\mu\delta B}{m_i}\right]^{1/2} = 4.09v_{ti0},
\end{eqnal}
and $\delta Q = Q(z=0)-Q(z=\Lz/2)$.
But our velocity grid only goes up to $\vparmax=3.75\,v_{ti0}$ causing these particles to run into our $\vpar$ boundary, producing an accumulation of $f_i$ (with a very large $\partial_{\vpar}f_i$) near the boundary due to the zero-flux BCs in velocity directions. This is the origin of the small bright spot near $(z,\vpar,\mu)=(\Lz/2,3.75v_{ti0},0.5776m_iv_{ti0}^2/(2\Bp))$ in figure~\ref{fig:adiabaticfi}(c).
This is not a numerical instability itself, because the maximum $f_i$ at the boundary grows only linearly in time, and it can be delayed significantly by increasing $\vparmax$ (at some computational cost).  This artifact is not noticeable in most of our simulations for the time duration that we ran them, and is worse here because of the large potential drop in the expander region for the Boltzmann electron case. But if the simulation continued for very long times ($\nu_{ee} t > 8/3$ for this Boltzmann electron case with these parameters), this numerical artifact can get so large that it triggers an instability of some sort, and the simulation quickly crashes.
This problem should be solved by a combination of improvements: a rigorous positivity-preserving scheme~\cite{Mandell2021thesis}, non-uniform velocity grids to more easily extend $\vparmax$, or perhaps most rigorously by extending $\vparmax$ to infinity and using exponential weighted basis functions there~\cite{Shi2017thesis}.
For the time being, we shall see in the next subsection that our force softening technique also has a beneficial effect in this regard. 



\subsubsection*{Force softening benchmark}

\begin{figure}
\includegraphics[width=0.48\textwidth]{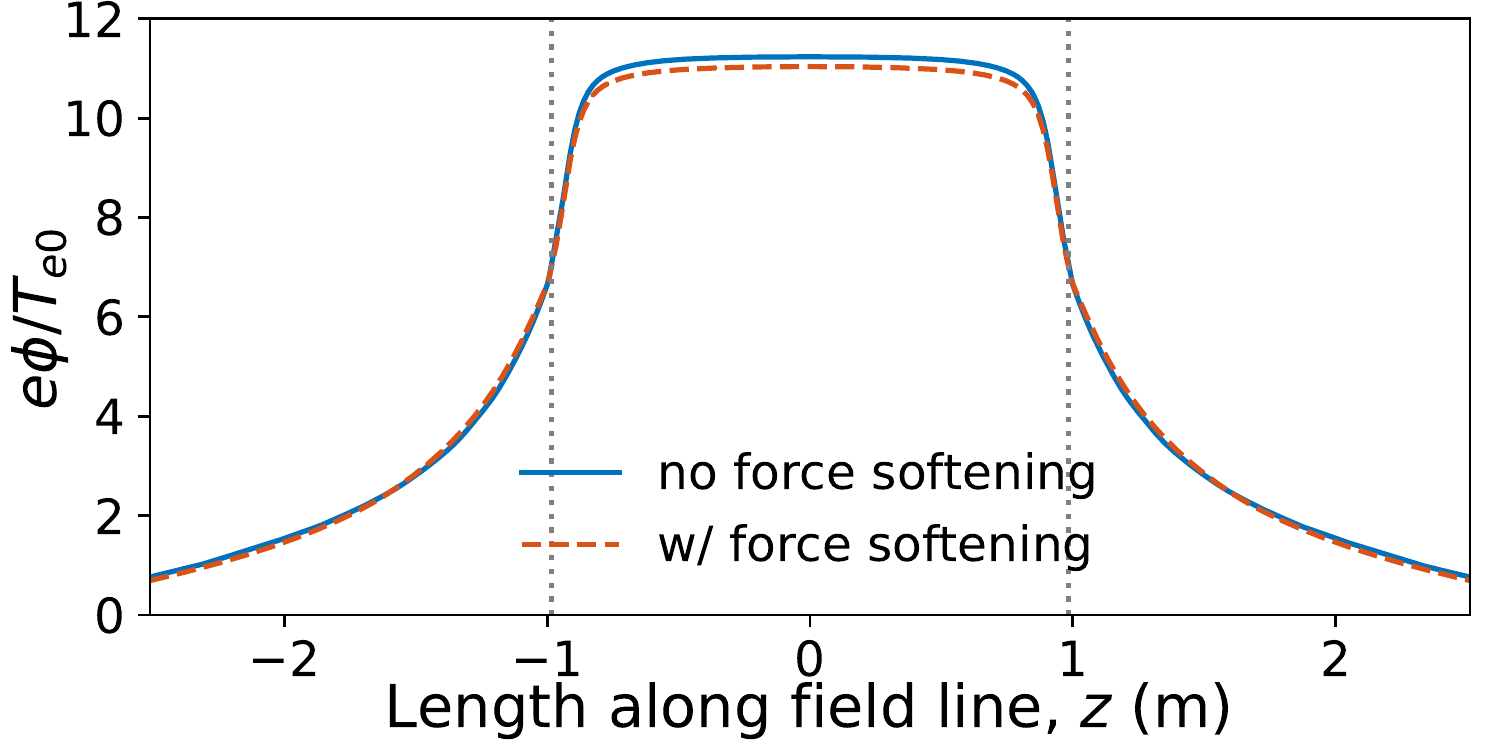}
\caption{\label{fig:phiForceSoftComp}Electrostatic potential at $\nu_{ee}t\approx8/3$ in Boltzmann electron simulations with (dashed orange) and without (solid blue) force softening. Note that the $\phi(z)$ variation is not accurate for $\abs{z}>\zm$ because Boltzmann assumptions break down in the expanders.}
\end{figure}

We carried out the same Boltzmann electron simulation employing the force softening technique described in section~\ref{sec:forcesoft} in order to overcome the CFL condition that makes $\Dt$ small. This strategy allowed us to run the same simulation up to $t=48\,\mu\mathrm{s}=8\nu_{ee}^{-1}/3$ in approximately 3.76 hours (on 288 cores), signifying a 19.5X speedup. When we compare the potential profile of these two simulations (figure~\ref{fig:phiForceSoftComp}) we find that the two agree very well. The small disagreement in the confined region is negligible considering that there are other shortcomings of our model that likely have a larger impact (e.g. collision model, neglect of beam ions, resolution, Boltzmann assumption).

Other quantities also exhibit agreement between the base Boltzmann electron simulation and that with force softening. For example, the ion number density and parallel fluid speed are very close in these two simulations, as shown in figures~\ref{fig:adiabaticMoms}(a)-(b) (compare solid and dotted-circled lines). Also, in figure~\ref{fig:adiabaticMoms}, we appreciate that force softening did not change the temperatures in the confined plasma much. It does lead to higher $T_{\parallel,\perp i}$ in the expander, particularly near the throats where $\Tperpi$ is significantly larger. Note that the perpendicular temperature in force softened simulations is calculated using the softened Hamiltonian, that is $T_{\perp s} = n_s^{-1}\int\mathrm{d}\v{w}\,\Pi(z,\mu)\mu B \BparS f_s$. These elevated temperatures can be a result of force softening allowing higher energy particles out to the expander than the experiment (or the simulation without force softening) does, which have a larger effect on the temperature than on the lower velocity moments. The density near the throats and in the expander also get very low, so the accuracy with which temperatures are computed is compromised. Nonuniform grids (see section~\ref{sec:compfuture}) will help us qualify these properties more accurately in subsequent studies.

\begin{figure}
\includegraphics[width=0.48\textwidth]{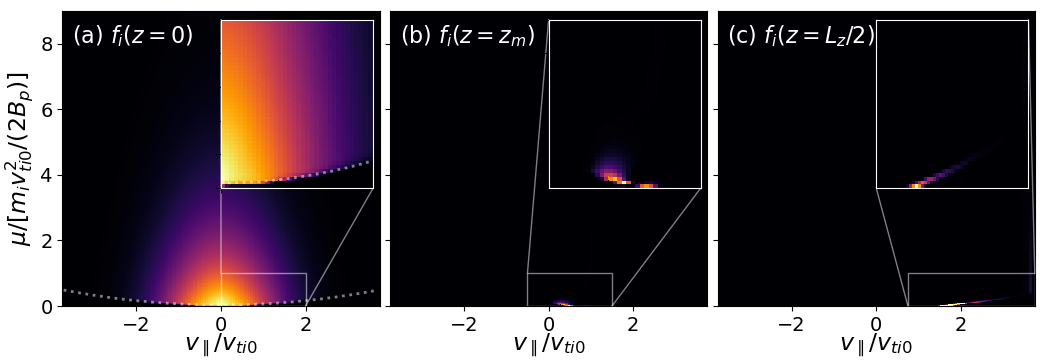}
\caption{\label{fig:forceSoftfi}Ion distribution function in Boltzmann electron simulation with force softening at $t=48\,\mu\mathrm{s}=8\nu_{ee}^{-1}/3$ at three locations along the field line: (a) the center, (b) the mirror throat, (c) the sheath entrance in the expander.}
\end{figure}

\begin{figure}
\includegraphics[width=0.48\textwidth]{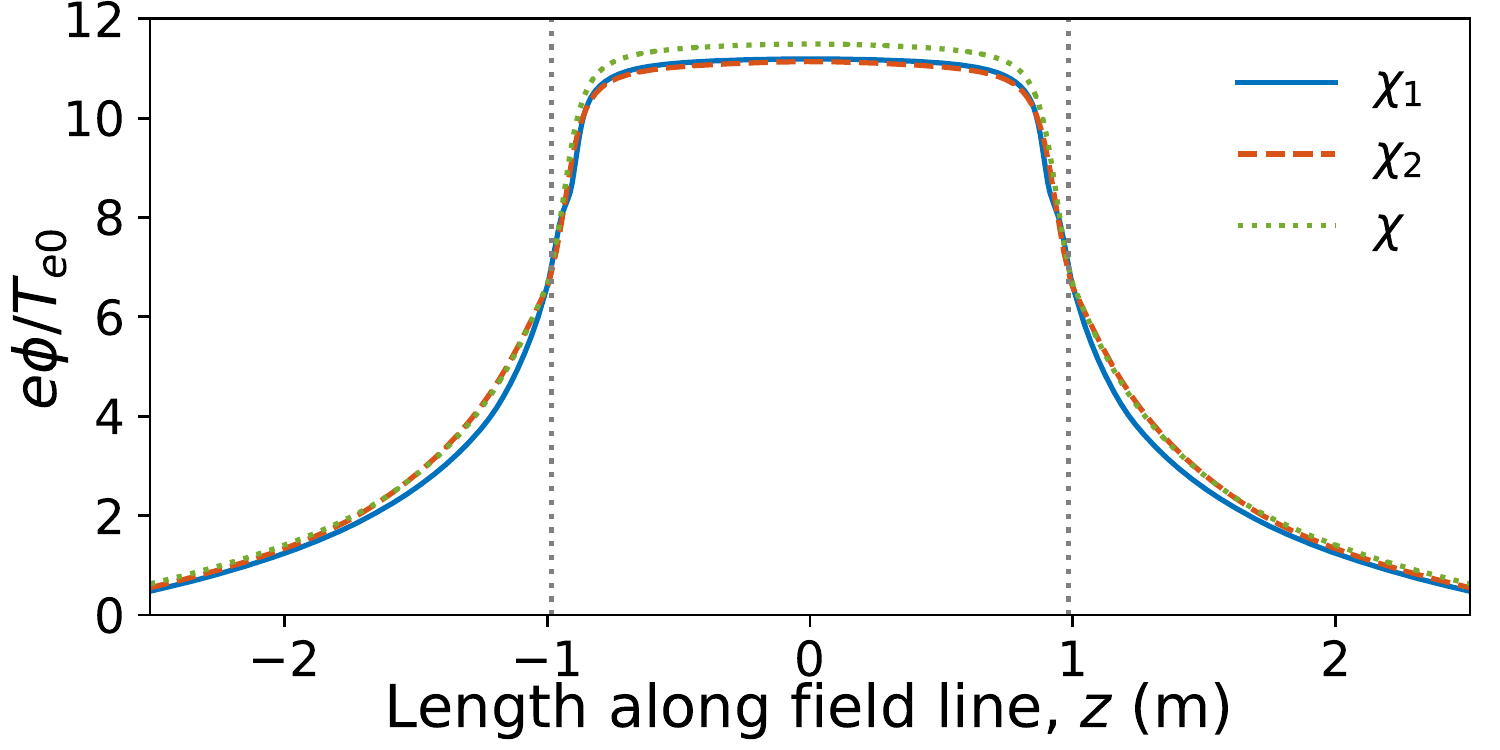}
\caption{\label{fig:phiForceSoftChiComp}Electrostatic potential at $\nu_{ee}t\approx5.56$ in Boltzmann electron simulations, using three types of softening factors (equations~\ref{eq:softChi} and~\ref{eq:softChiAlternatives}).}
\end{figure}

Good qualitative agreement between simulations with and without force softening is seen in the ion distribution function, shown in figure~\ref{fig:forceSoftfi} at the same three locations as in figure~\ref{fig:adiabaticfi}. The most notable difference is that the numerical anomaly previously present near $(z,\vpar,\mu)=(\Lz/2,3.75v_{ti0},0.5776m_iv_{ti0}^2/(2\Bp))$ is now absent. 
Now that the distribution function is monotonically decaying near the velocity boundary and that the simulation advances faster than without force softening, we can time integrate for longer periods of time. Moreover, we performed the simulation to $t=100\,\mu\mathrm{s}\approx5.56\nu_{ee}^{-1}$ with two other softening factors,
\begin{eqnal} \label{eq:softChiAlternatives}
    \chi_1(z,\mu) &= \begin{cases}
        1 \qquad &\mu\leq\mu_t(z), \\
        \mu_t(z)/\mu \qquad &\mu>\mu_t(z),
    \end{cases} \\
    \chi_2(z,\mu) &= \begin{cases}
        1 \qquad &\mu\leq\mu_t(z), \\
        \frac{1}{2}\left[1+\tanh\left(\frac{\mu_t(z)-\mu}{\mu_t(z)/20}\right)\right] \qquad &\mu>\mu_t(z),
    \end{cases}
\end{eqnal}
in order to assess how sensitive the result may be to this choice. We see in figure~\ref{fig:phiForceSoftChiComp} that for the three softening factors used, the resulting potential is almost the same in each case. Additionally, note by comparing figures~\ref{fig:phiForceSoftComp} and~\ref{fig:phiForceSoftChiComp} that the profiles have changed little from $\nu_{ee}t=8/3$ to $\nu_{ee}t=5.56$; most of the potential profile in Boltzmann electron simulations is built within a couple of electron collision periods.

Another important thing to note about this potential profile is that the potential drop between the plasma center and the mirror throat is $e\Delta\phi/T_{e0}=4.47$. This indicates that there is strong electrostatic confinement of the electrons and that the potential drop is close to yet exceeds the Pastukhov estimate of $e\Delta\phi/T_{e0}=3.72$ in figure~\ref{fig:pastukhovPhi}. One possible explanation for the increased potential drop is that the Boltzmann approximation essentially implies highly collisional electrons, for which the Pastukhov calculation can yield potentials above $\simeq e\Delta\phi/T_{e0}=4.47$ (see figure~\ref{fig:pastukhovPhi}(b)). It is also apparent that according to this model, there is an additional, strong potential drop in the expander of $e(\phi(z=\zm)-\phi(z=\pm\Lz/2))/T_{e0}=6.39$ that further confines the electrons. As mentioned earlier the Pastukhov calculation is not valid in the expander, so this Boltzmann electron simulation is a first rough estimate of what the potential profile may be in that region. At the same time, we know that the isothermal, Boltzmann electron assumption is invalid in the expander since we would expect large temperature drops, relatively collisionless plasma (due to the rapid drop in density), and non-Maxwellian distributions~\cite{Ryutov2005} therein. Other recent semi-analytic estimates of this potential structure actually produce little to no potential drop in the expander~\cite{Egedal2022}. More accurate estimations of the profile may be attained by relaxing the isothermal, Boltzmann assumption or using kinetic electrons.

\subsection{Kinetic electrons} \label{sec:resultsKinElc}

The kinetic electron model evolves the gyrokinetic equation~\ref{eq:gkeq1d} for both ions and electrons, and computes the electrostatic potential using a Poisson equation with either a linearized polarization (equation~\ref{eq:poisson1dlinear}) or a nonlinear polarization (equation~\ref{eq:poisson1dnonlinear}); we present results for both field models in this section.

\subsubsection*{Linear polarization}

\begin{figure}
\includegraphics[width=0.48\textwidth]{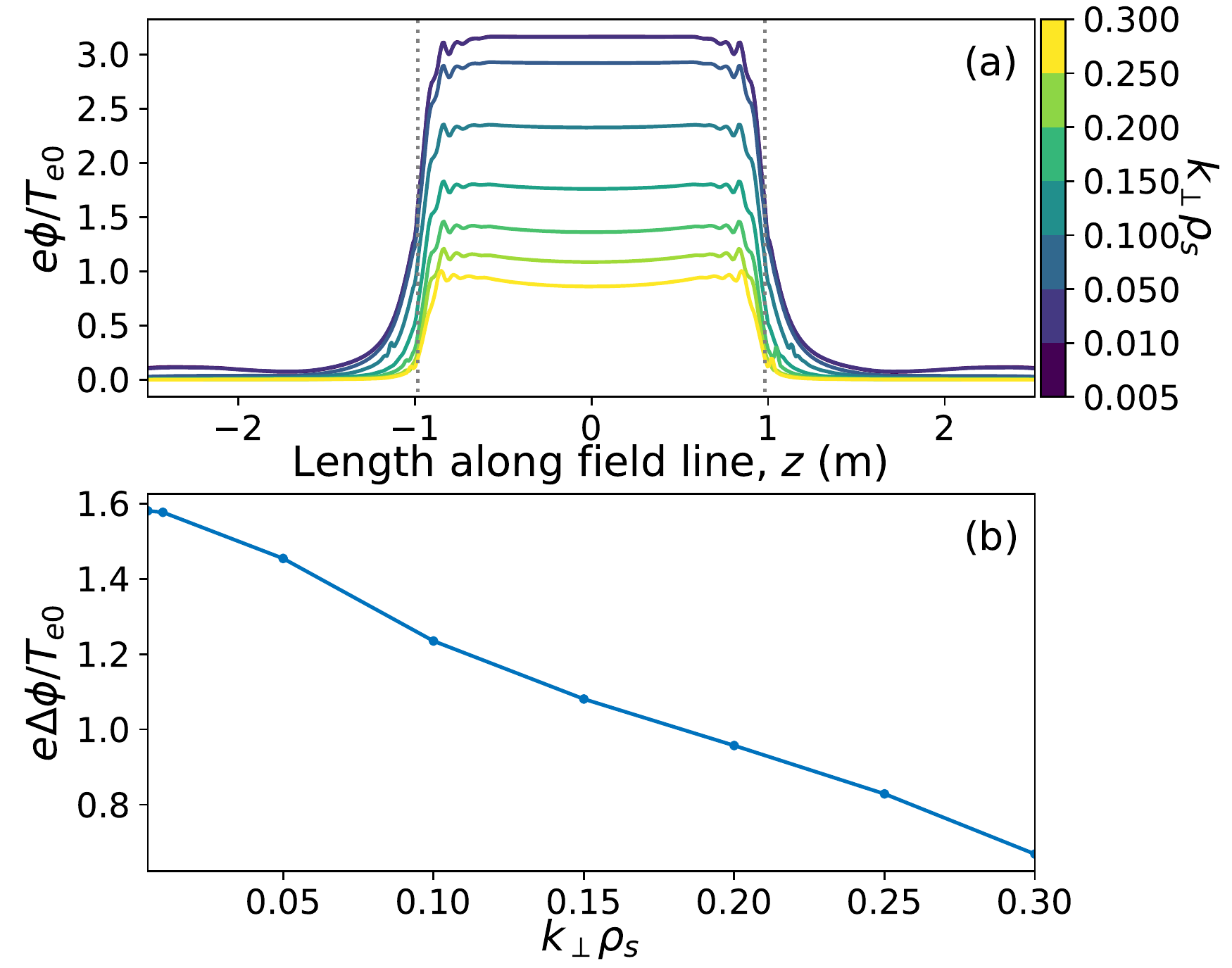}
\caption{\label{fig:phikperpScan}(a) Electrostatic potential profiles in kinetic electron simulations with linearized polarization, and (b) the potential drop from the center ($z=0$) to the mirror throat ($z=\zm$) normalized to $T_{e0}/e$ at $t=8\,\mu\mathrm{s}$ as a function of $\kperp\rho_s$.}
\end{figure}

Kinetic electron simulations with a linear polarization were performed with the $288\times64\times192$ grid mentioned in section~\ref{sec:comp}, which was deemed the most suitable according to the resolution scan in appendix~\ref{sec:resconv}. These simulations require the choice of the parameter $\kperp$ in equation~\ref{eq:poisson1dlinear}, and thus one of our first tasks was to examine the effect this parameter has on the resulting potential. In principle, the lowest possible value of $\kperp\rhos$ would be preferable in order to approximate the $\kperp\rhos\to0$ limit we are interested in, given our choice of 1D models. Additionally, as we would expect from equation~\ref{eq:poisson1dlinear}, smaller $\kperp\rhos$ would lead to larger potentials, possibly as large as those seen in previous experiments or the Pastukhov potential estimate at the beginning of this section. Such an increase in $\phi$ with decreasing $\kperp\rhos$ is precisely what we see in figure~\ref{fig:phikperpScan}(a), showing the potential at $t=8\,\mu\mathrm{s}$. Precisely how fast the potential increases with $(\kperp\rhos)^{-1}$ can be seen in figure~\ref{fig:phikperpScan}(b), where we provide the potential drop between the center and the mirror throats ($\Delta\phi$) for each $\kperp\rhos$. Extrapolating to $\kperp\rhos=0$ or examining the points at $\kperp\rhos=\{0.005,\,0.01\}$ on may conclude that the asymptotic limit has been reached by $\kperp\rhos=0.01$, although it is difficult to claim that with certainty at this point. We would have liked to continue this study at lower $\kperp\rhos$, but as explained in section~\ref{sec:dtconstraints} the $\omega_H$ frequency increases with $(\kperp\rhos)^{-1}$ and therefore makes the simulations impractically expensive for $\kperp\rho_s<0.005$. One can also estimate that in \wham~(with $\Bp=0.86,~\Rm=20$) $\rho_s=5.17\times10^{-3}\,\mathrm{m}$, and the smallest perpendicular mode that fits in a $2\alim$-diameter plasma is $\kperp\approx\pi/(2\alim)$ such that $\kperp\rhos=0.06$. If one further considers that $B(z)\,a(z)^2$ is a constant along $z$, with $a(z)$ being the minor radius of the plasma, $\kperp\rhos$ can be 0.015 (keeping temperature constant) or smaller near the mirror throat.



\begin{figure}
\includegraphics[width=0.48\textwidth]{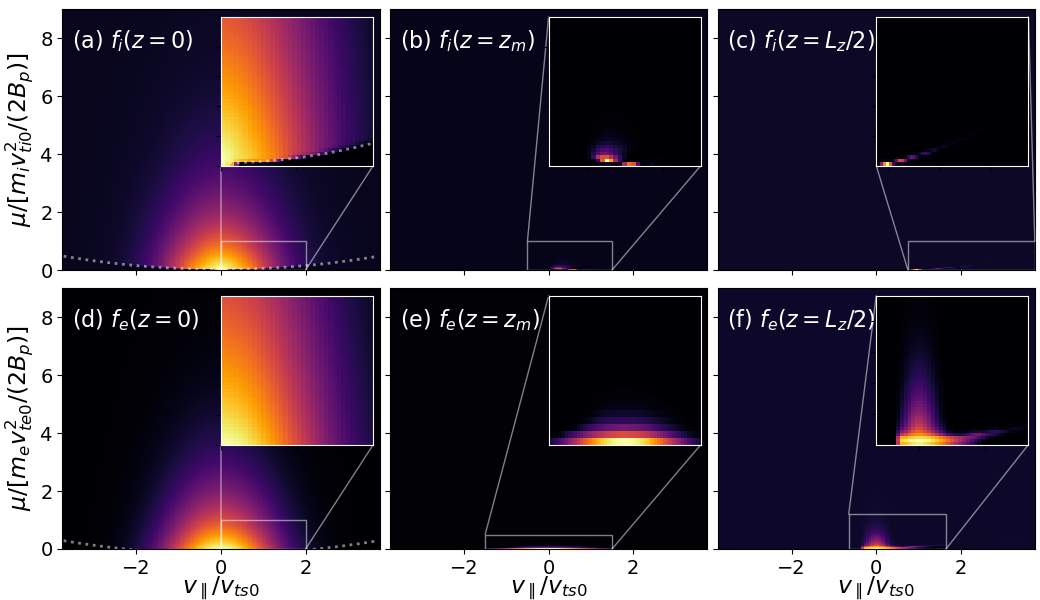}
\caption{\label{fig:kineticf}Ion (top row) and electron (bottom row) distribution functions in kinetic electron simulation with linearized polarization at $t=32\,\mu\mathrm{s}\approx1.78\nu_{ee}^{-1}$ and three locations along the field line: (a,d) the center, (b,e) the mirror throat, (c,f) the sheath entrance in the expander.}
\end{figure}

Based on such $\kperp\rho_s$ scan, we decided to time integrate the kinetic electron simulation with $\kperp\rho_s=0.01$ for a longer period of time. By employing 288 cores for 1,019 hours, this kinetic electron simulation was advanced to $t=32\,\mu\mathrm{s}\approx1.78\nu_{ee}^{-1}$, at which point the ion and electron distribution functions at the center, mirror throat and sheath entrance looked as shown in figure~\ref{fig:kineticf}(a-c). Through comparison with figure~\ref{fig:adiabaticfi} we see that the structure of $f_i$ is similar to that in the Boltzmann electron simulations. The ion distribution function obeys the expected trapped passing boundary at $z=0$, and has a slanted arrangement at the sheath entrance caused by the decreasing Yushmanov potential. It is notable, however, that in this case, we do not see an accumulation of $f_i$ near the $\vpar$ boundary as we did in the Boltzmann electron case. That is because, as we show below, the potential drop in the expander of the kinetic electron model is very small and leads to a much weaker electrostatic acceleration of the ions in the expander.

A kinetic electron simulation also gives us insight into the phase space structure of the electrons, and we provide a snapshot of it in figure~\ref{fig:kineticf}(d-f). A first observation of this $f_e$ makes it apparent that although electrons seem Maxwellian in the confined plasma, consistent with the Boltzmann model, but as one moves into the expander and approaches the sheath entrance the distribution acquires non-Boltzmann features. This is direct evidence of the limitations of the Boltzmann model, and may also help us derive better reduced electron models. The electron distribution function in the mirror center ($z=0$) also obeys the expected trapped-passing boundary given in equation~\ref{eq:trappedpassing}. However, it can be difficult to discern in figure~\ref{fig:kineticf}(d) for three reasons. First, the electrostatic force pushes these boundaries away from the origin so that some particles with $\mu=0$ (and no mirror confinement) are still confined. Second, in $\vpar-\mu$ space, the loss cone is actually a parabola, making it harder to discern the familiar cone. The HTS coils and the electrostatic potential also provide such good electron confinement that the loss cones are very narrow and far from $\vpar=0$. And third, where the loss cones appear in this $\vpar-\mu$ space, $f_e$ is so small that it is difficult to appreciate its details (unless one plots in other ways and sacrifices visualizing other features). In the expander (figure~\ref{fig:kineticf}(f)) we also see some signs of Yushmanov acceleration (the branch of $f_e$ that extends towards higher $\vpar$ and $\mu$). And in contrast to the ions, the electrons have a population that propagates back towards the machine (i.e. $f_e(\vpar<0)$), which is a result of the reflection by the sheath potential and is truncated at $\vpar\approx\left(2e\phi(z=\Lz/2)/m_e\right)^{1/2}$ as we would expect from our sheath BCs~\cite{Shi2017}. 

\begin{figure}
\includegraphics[width=0.48\textwidth]{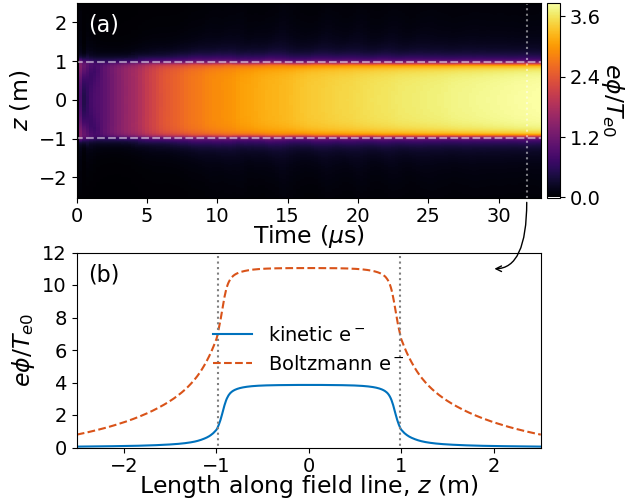}
\caption{\label{fig:kinPhi}(a) Electrostatic potential $\phi(z,t)$ in simulation with kinetic electrons and linearized polarization (mirror throats indicated with white dashed lines), and (b) electrostatic potential profile at $t=32\,\mu\mathrm{s}\approx1.78\nu_{ee}^{-1}$. Note that the $\phi(z)$ variation in the Boltzmann electron case is not accurate for $\abs{z}>\zm$ because Boltzmann assumptions break down in the expanders.}
\end{figure}

The sheath potential at this time, $e\phi(z=\pm\Lz/2)/T_{e0}=0.0552$ is considerably smaller than the $e\phi(z=\pm\Lz/2)/T_{e0}=0.7924$ produced by the Boltzmann electron simulation (see figure~\ref{fig:kinPhi}(b)). This is partly because the latter results from ambipolar sheath fluxes and the Boltzmann approximation, while the former is established by the separation of ion and electron guiding center charge density in the Poisson equation. The difference between Boltzmann and kinetic electron particle dynamics in the expander, which sets the potential therein, is also partly due to the short electron transit time that precludes thermalization in the kinetic electron case (see section~\ref{sec:adiabatic}). In the Boltzmann electron case the electrons are already thermalized, while with kinetic electrons one must evolve the system for at least $\geq\nu_{ee}^{-1}$ to give electrons a chance at thermalization. Therefore, comparisons between Boltzmann and kinetic electron results should be viewed keeping in mind that the latter may not have run long enough to reach steady state or thermalized electrons. Such limitation is also apparent in figure~\ref{fig:kinPhi}(a), which shows that by the end of this simulation near $\nu_{ee}t=1.83$ the electrostatic potential continues to rise. When we analyze the potential profiles away from the sheath, we see that the potential drop between the center and the mirror throats ($e\Delta\phi/T_{e0}\approx2.57$) is lower than that obtained with Boltzmann electrons ($e\Delta\phi/T_{e0}\approx3.94$). Furthermore, there is a large potential drop in the expander seen with Boltzmann electrons that is absent with kinetic electrons. Both of these facts could be partially explained by the large $\nu_{ee}$ implied by the Boltzmann assumption, which would cause additional potential drops due to the stronger electron scattering and thus charge separation that would require a larger potential to maintain quasineutrality. As discussed in section~\ref{sec:adiabatic} we do not expect the Boltzmann model to be accurate in the expander, though, and is simply an intermediate tool that may be honed in future work.

\begin{figure}
\includegraphics[width=0.48\textwidth]{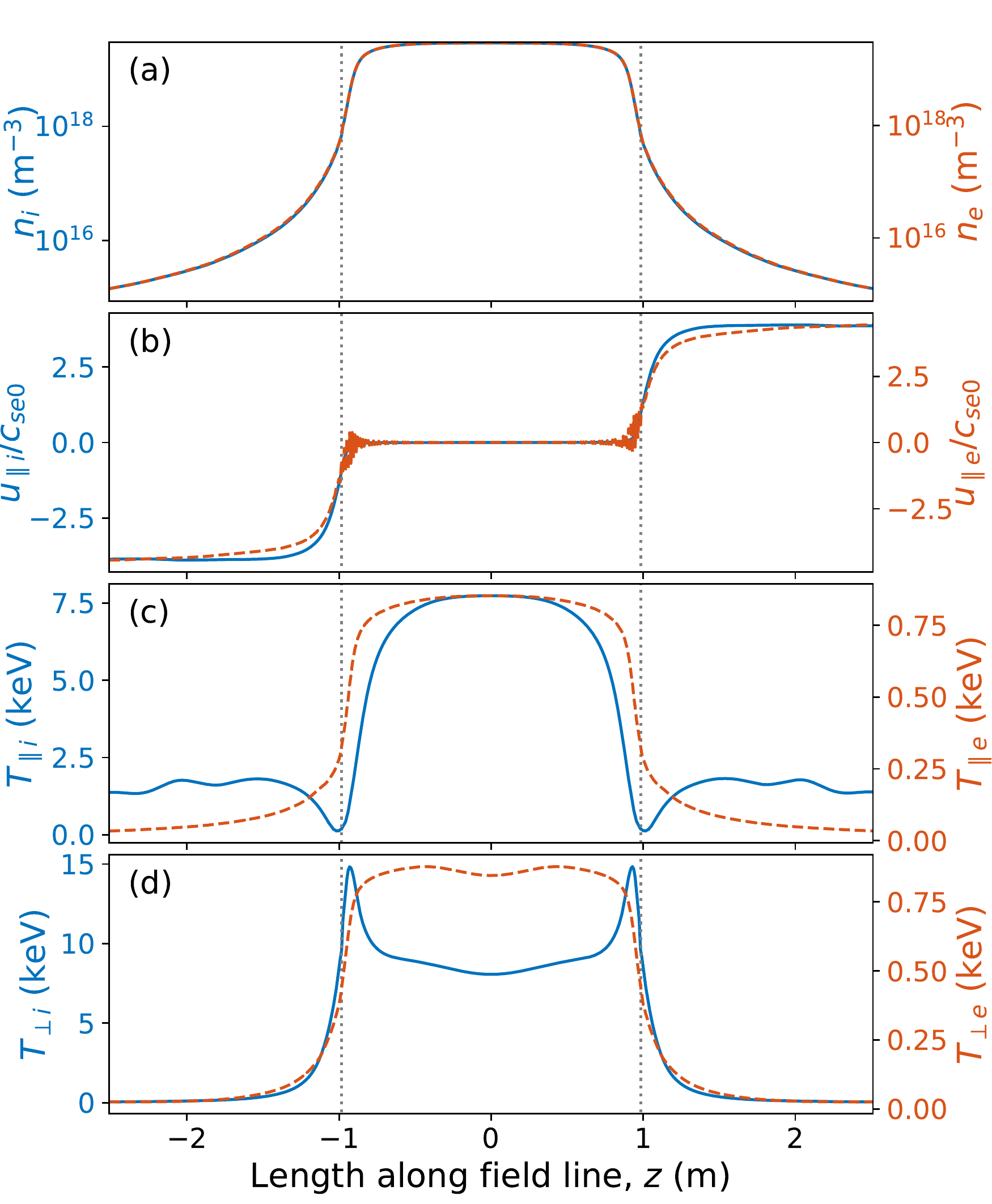}
\caption{\label{fig:kineticMoms} Velocity moments of the ion (solid blue) and electron (dashed orange) distribution functions at $t=32\,\mu\mathrm{s}\approx1.78\nu_{ee}^{-1}$ in kinetic electron simulation with linearized polarization.}
\end{figure}

It is also interesting to examine the velocity moments of this kinetic electron model, provided in figure~\ref{fig:kineticMoms}. Although given at a slightly different time, these ion moments are similar to those in the Boltzmann electron simulation (figure~\ref{fig:adiabaticMoms}). The main difference is the lower mean flow speed ions are accelerated to in the expander of the kinetic electron simulation, $\upari\approx\pm4c_{se0}$ vs. $\upari\approx\pm5.5c_{se0}$ with Boltzmann electrons. Ion densities and temperatures are nearly the same in both cases. Additionally, we can also see the resulting electron moments now. The first two of these moments, $n_e$ and $\upare$, resemble their ion counterparts. The electron temperatures, however, are much lower and similar to each other, indicating much higher levels of isotropy than it is found for the ions. This is to be expected based on the large $\nu_{ee}/\nu_{ii}$ ratio, the fact that we have evolved the system for $\nu_{ee}t>1$ but $\nu_{ii}t\ll1$, and previous theory and simulation yielding essentially Maxwellian electrons in the confined plasma~\cite{Post1987}.

\subsubsection*{Nonlinear polarization and force softening}

The linear polarization model did not allow time integration for longer periods because, even if one used force softening, the $\omega_H$ mode frequency was very large for $\kperp\rho_s=0.01$ and made the time step very small. Our second alternative to explore potential formation with kinetic electrons is to use the nonlinear polarization model (equation~\ref{eq:poisson1dnonlinear}) with force softening. At first, we attempted to use the same $288\times64\times192$ grid employed for previous Boltzmann and kinetic electron simulations, reaching $t=32\,\mu\mathrm{s}\approx1.78\nu_{ee}^{-1}$ with 288 cores in 107.72 hours (a >9X improvement in performance). However, shortly after this phase the simulation became unstable, possibly due to insufficient resolution near the mirror throat and in the expander. We therefore carried out a second kinetic electron simulation with nonlinear polarization and force softening but using a $288\times96\times192$ grid, i.e. $\Nvpar$ was increased by 50\%. This section focuses on this second, higher resolution simulation.

\begin{figure}
\includegraphics[width=0.48\textwidth]{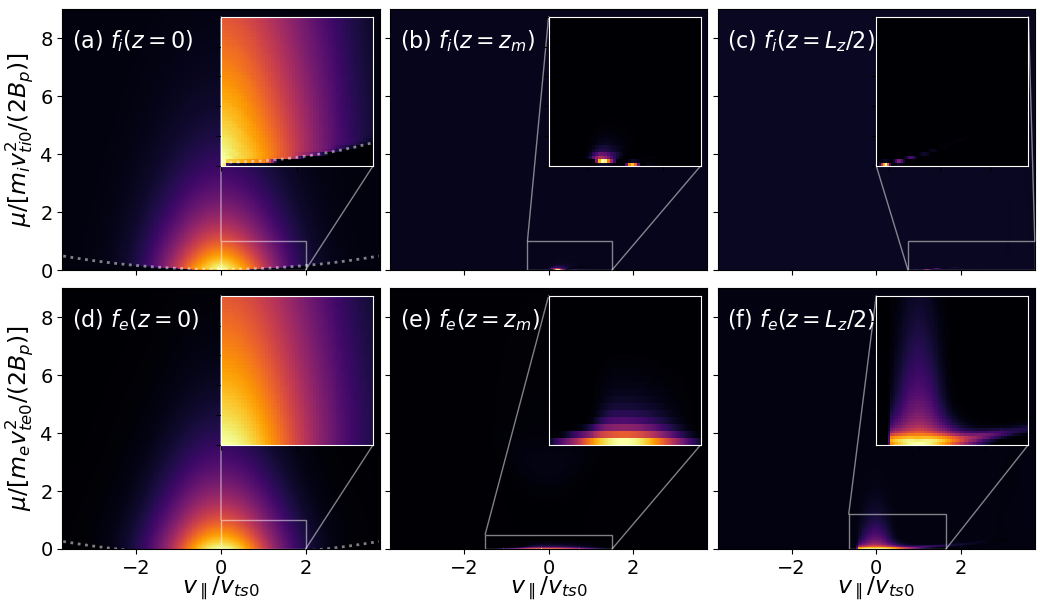}
\caption{\label{fig:nonlinPolf}Ion (top row) and electron (bottom row) distribution functions in kinetic electron simulation with nonlinear polarization at $t=32\,\mu\mathrm{s}\approx1.78\nu_{ee}^{-1}$ and three locations along the field line: (a,d) the center, (b,e) the mirror throat, (c,f) the sheath entrance in the expander.}
\end{figure}

\begin{figure}
\includegraphics[width=0.48\textwidth]{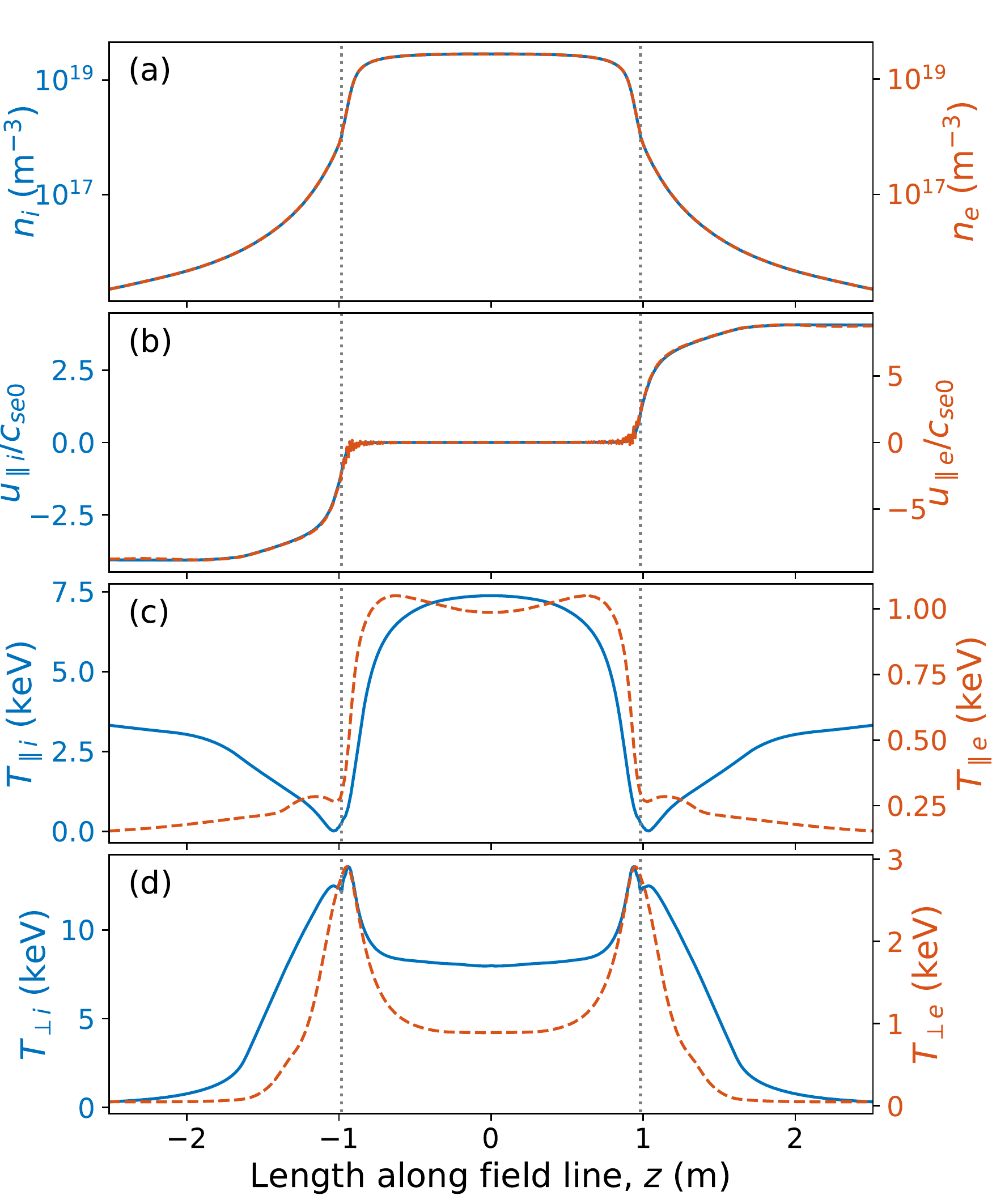}
\caption{\label{fig:nonlinMoms} Velocity moments of the ion (solid blue) and electron (dashed orange) distribution functions at $t=32\,\mu\mathrm{s}\approx1.78\nu_{ee}^{-1}$ in kinetic electron simulation with nonlinear polarization and force softening.}
\end{figure}

The kinetic electron simulation with nonlinear polarization reached $t=32\,\mu\mathrm{s}\approx1.78\nu_{ee}^{-1}$ in approximately 213 hours on 288 cores. At this time the distribution functions at the center, mirror throat and sheath entrance looked very similar to those in the simulation with linearized polarization (compare figures~\ref{fig:kineticf} and~\ref{fig:nonlinPolf}). The distributions at the center of the device exhibit the expected loss paraboloids (dotted white lines in figures~\ref{fig:nonlinPolf}(a,d)), and in the sheath entrance $f_e$ is truncated at the cutoff velocity corresponding to the sheath potential. Although hard to appreciate, this cutoff is at a lower $\vpar$ in figure~\ref{fig:kineticf}(f) than in figure~\ref{fig:nonlinPolf}(f), suggesting that the sheath potential is higher in the nonlinear polarization simulation. More noticeable differences are seen in the velocity moments, at least between kinetic electron simulations with linear (figure~\ref{fig:kineticMoms}) and nonlinear polarization (figure~\ref{fig:nonlinMoms}). There is slightly more separation between the guiding center densities in the nonlinear polarization case than in the linear polarization, although the logarithmic scales obscure it; such separation foretells a taller potential profile, as shown below. The ion parallel mean speed raises to the same amplitude in both simulations near or past the mirror throat, but in both cases, it is inferior to its value in the Boltzmann electron case (figure~\ref{fig:adiabaticfi}). On the other hand, $\upare$ saw a considerable increase in the nonlinear polarization model. Focusing on the temperatures next, we observe that although some differences appear in the tails of $\Tpari$ (solid blue in figure~\ref{fig:nonlinMoms}(c)), the profile is nearly identical to that obtained in the Boltzmann electron simulation with force softening (dotted line with circles in figure~\ref{fig:adiabaticMoms}(c)). Likewise, the $\Tperpi$ profile in figure~\ref{fig:nonlinMoms}(d) is different in that it is augmented in the near-expander compared to the linearized polarization case (figure~\ref{fig:kineticMoms}(d)), but the former is likely affected by force softening since such profiles follow the same trends as those in dotted-circle blue lines in figure~\ref{fig:adiabaticMoms}.

\begin{figure}
\includegraphics[width=0.48\textwidth]{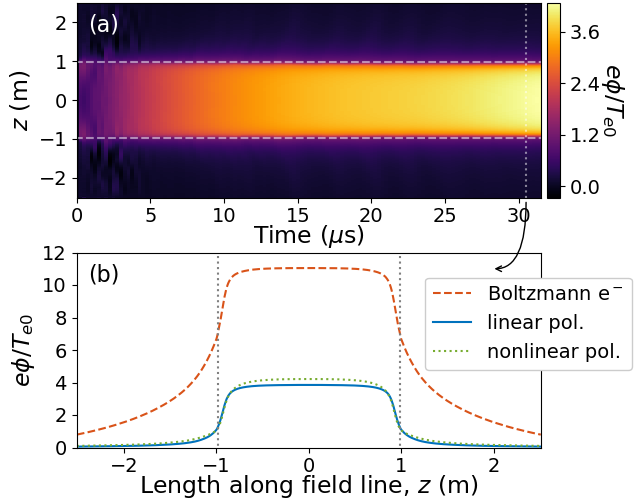}
\caption{\label{fig:nonlinPhi}(a) Electrostatic potential $\phi(z,t)$ in simulation with kinetic electrons, nonlinear polarization and force softening (mirror throats indicated with white dashed lines), and (b) electrostatic potential profile at $t=32\,\mu\mathrm{s}\approx1.78\nu_{ee}^{-1}$, with the profiles from Boltzmann electron and linearized polarization kinetic electron simulations given for comparison. Note that the $\phi(z)$ variation in the Boltzmann electron case is not accurate for $\abs{z}>\zm$ because Boltzmann assumptions break down in the expanders.}
\end{figure}

The difference in guiding center densities mentioned above is only responsible for a slight increase in the height of the potential profile, as shown by solid blue and green dotted lines in figure~\ref{fig:nonlinPhi}. At this time, the linear and nonlinear polarization simulations produce nearly identical profiles. In that same figure, we also compare the $\phi$ profiles with that obtained by the Boltzmann electron simulation. The potential drop between the center and the throat is somewhat similar in the Boltzmann electron and nonlinear polarization simulations, $3.87T_{e0}/e$ and 
$3.09T_{e0}/e$ (or $3.12T_e(z=0)/e$), respectively. The largest difference is in the value of the sheath potential and the potential drop seen in the expander ($\abs{z}>z_m=0.98\,\mathrm{m}$). The absence of a potential drop in the expander is unlike what is seen in some GDT experiments, where $\phi(z=\zm)-\phi(z=\Lz/2)$ carries most of the drop across the device depending on the expansion ratio of the magnetic field therein ($K_w=\Bm/B(\Lz/2)$)~\cite{Ivanov2013}. The Boltzmann electron model may be more accurate for the more collisional GDT, but is not suitable for the large $K_w$ and collisionless regimes in \wham~\cite{Ryutov2005}, as discussed in section~\ref{sec:adiabatic}. Therefore the lack of a potential drop in the expander of kinetic electron simulations is not necessarily incorrect, although at this time ($t=32\,\mu\mathrm{s}=1.78\nu_{ee}^{-1}$) the profiles have not stopped evolving, as can be appreciated from figure~\ref{fig:nonlinPhi}(a). And even in the recent work that estimated \wham's $\phi(z)$ using semi-analytical methods~\cite{Egedal2022}, no potential drop was seen in the expander.

\subsection{Ambipolar potential development} \label{sec:phidev}

One of the beneifts of the Boltzmann electron model and kinetic electron models with force softening is the ability to run them for a much longer time due to their lower computational cost. This allows better assessments of what the steady state profiles look like and better approximations of an equilibrium that could be used for 3D turbulence calculations in the future. We thus time-integrated these two models up to $t=100\,\mu\mathrm{s}\approx5.56\nu_{ee}^{-1}$. As noted in the previous section, the resolution along $\vpar$ had to be increased in the nonlinear polarization simulation, bringing the cost of this longer simulation to 749.8 hours on 288 cores. And to make a fairer comparison, we ran a second Boltzmann electron simulation with the same $288\times96\times192$ resolution as the nonlinear polarization simulation for 14.73 hours on 288 cores (a little less than a factor of $\sqrt{m_i/m_e}\approx60$ difference).

\begin{figure}
\includegraphics[width=0.48\textwidth]{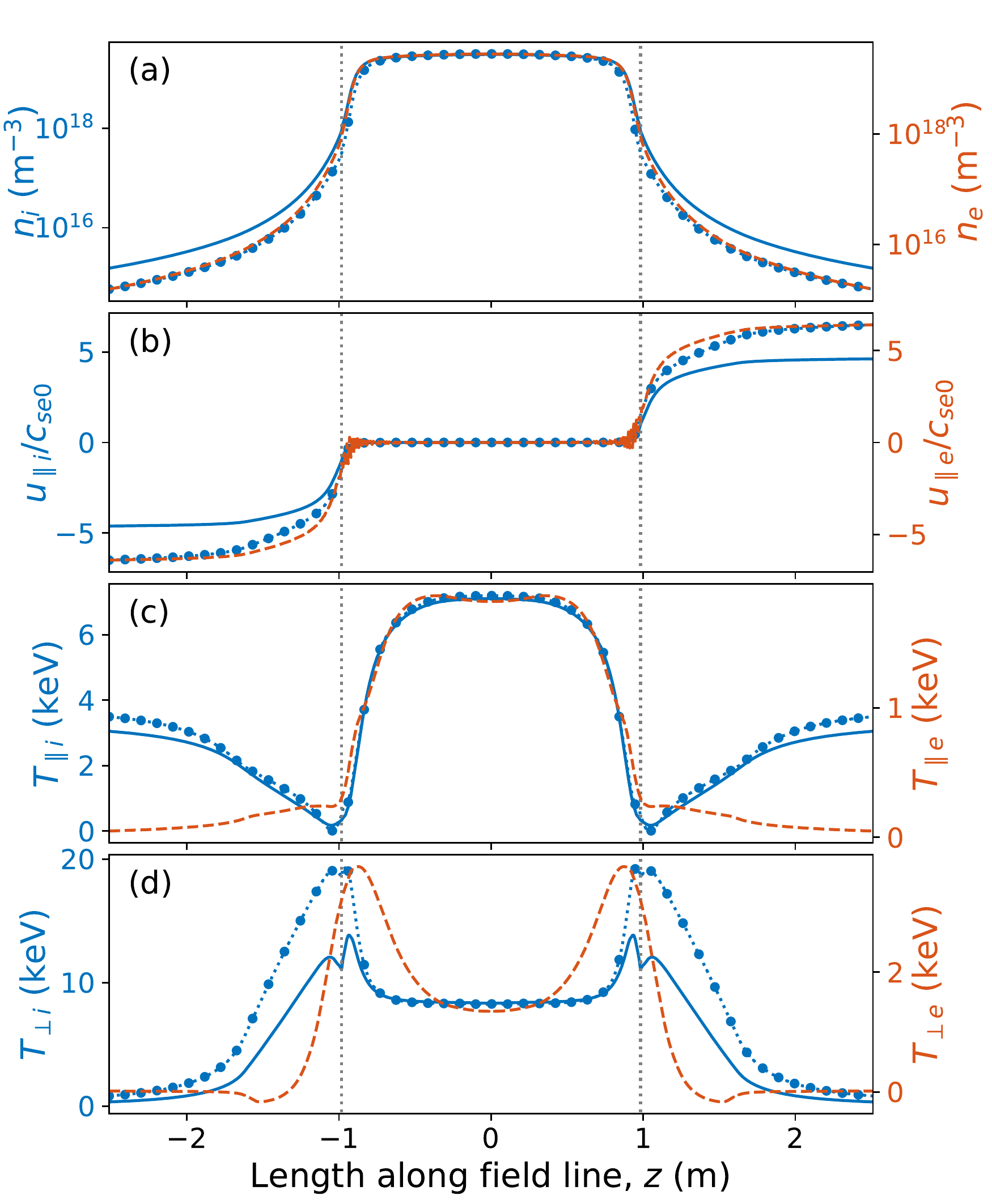}
\caption{\label{fig:momComp} Guiding center density ($n_s$), mean drift speed ($u_{\parallel s}$) normalized to $c_{se0}=\sqrt{T_{e0}/m_i}$, and parallel and perpendicular temperatures ($T_{\parallel s}$, $T_{\perp s}$) in kinetic electron simulation with nonlinear polarization (solid blue, dashed orange), and Boltzmann electron simulation (dotted blue with circles) at $t=100\,\mu\mathrm{s}\approx5.56\nu_{ee}^{-1}$, both with force softening. Dotted vertical lines indicate mirror throats.}
\end{figure}

These two simulations developed similar ion guiding center density and parallel temperature profiles (solid blue and dotted-circled blue in figures~\ref{fig:momComp}(a,c)), aside from a larger drop in the expander $n_i$ of the Boltzmann electron simulation. Additional quantitative differences are found in the speed ions accelerate to within the expander, and in $\Tperpi$ near the mirror throats which as explained below may be affected by force softening. One of the most contrasting results is the large $\Tperpe/\Tpare$ ratio near the mirror throats given that, owing to their frequent collisions, we would expect the electron $f_s$ to be close to isotropic. Near the center of the plasma, this ratio, $1.35/1.82\approx0.74$, is not far from unity, but at the mirror throat, it rises to $3.76/0.9\approx4.18$. It is possible, although it remains to be shown, that since the structure of $f_e$ in $z-\mu$ is different to $f_i$ the force softening function $\chi$ that does not incorporate the effect of the electrostatic potential is causing some of this modification of $\Tperpe$ near the mirror throat. Future numerical experiments are planned to explore this possibility.

\begin{figure}
\includegraphics[width=0.48\textwidth]{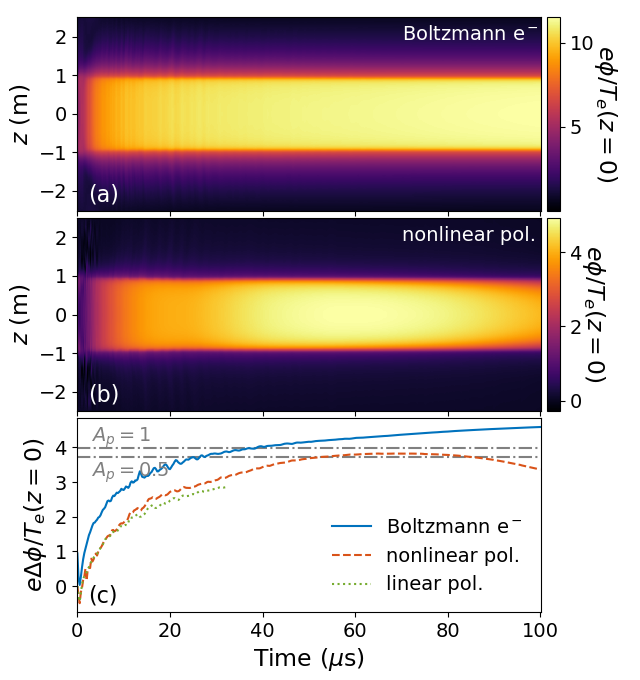}
\caption{\label{fig:phiComp}Spatio-temporal evolution of the electrostatic potential in Boltzmann electron simulation (top), and kinetic electron simulation with nonlinear polarization (middle). The potential drop between the mirror center and throats ($\Delta\phi$) normalized by the central electron temperature ($T_e(z=0)$) is shown as a function of time (bottom). The $\Delta\phi$ estimated from the Pastukhov loss rates are indicated with horizontal grey dash-dot lines.}
\end{figure}

Another noticeable feature of the electron temperatures in figure~\ref{fig:momComp}(e,d) is that the scalar temperature, $T_e=(\Tpare+2\Tperpe)/3$, has risen considerably over the $T_{e0}=940\,\mathrm{eV}$ reference temperature. And this phenomenon has an impact on a vital metric of this study, quantifying whether the ambipolar potential developed is strong enough to confine the electrons. Figure~\ref{fig:phiComp} shows the evolution of the $\phi$ profiles for the Boltzmann and nonlinear polarization simulations. The bottom panel provides the potential drop between the center and the throats normalized to the local electron temperature. The first thing to notice is that the Boltzmann electron model surpasses the semi-analytic estimate of $e\Delta\phi/T_{e0}=3.98$ obtained at the beginning of section~\ref{sec:results}. As discussed previously, Boltzmann electrons imply high collisionality and can thus lead to larger potential drops, hence why the solid blue line surpasses the grey dash-dot $A_p=1$ line in figure~\ref{fig:phiComp}(c). On the other hand, the nonlinear polarization simulation (dashed orange line) also attained a $\Delta\phi$ comparable to the $3.72T_{e0}/e$ estimate obtained earlier (with $A_p=0.5$) by $t\approx50\,\mu\mathrm{s}\approx2.78\nu_{ee}^{-1}$. However, we see this curve downturn later and dip below this semi-analytic estimate. Although it is still within the expected range of several $T_e/e$, this dip is because the electron temperature in the nonlinear polarization simulation continues to increase at this time. If we instead had normalized $e\Delta\phi$ to $T_{e0}$ in figure~\ref{fig:phiComp}, we would have seen the orange dashed curve follow a similar trend as the blue curve for the Boltzmann electron simulation. We also include the potential drop in the linear polarization kinetic electron simulation in dotted green, which was not time-integrated as long due to its increased computational cost but seems to follow a similar trend to the nonlinear polarization model. The tentative conclusion here is that both the Boltzmann electron and the kinetic electron models develop a potential profile with sufficient height to confine the electrons, with the caveat that the kinetic electron simulation continues to heat, an aspect that will be investigated in detail in future work.

\section{Summary} \label{sec:conclusion}

In this work we aimed to understand the computational challenges of studying high-field mirrors with large mirror ratios using a continuum gyrokinetic model, and to determine if the model used produces the approximate (Pastukhov) ambipolar potential ($\phi$) necessary to confine the electrons. High accuracy estimations of $\phi$ require longer simulations (e.g. several ion collision times) for which bounce-averaged Fokker-Planck or implicit methods are better suited, so that particle loss rates reach steady state in the presence of a self-consistent potential. In spite of that, the ambipolar potential only depends on the ion loss rate logarithmically, so simulations done for a few electron collision times, as done here, adequately verify if the main physics of the ambipolar potential formation is handled moderately well. Our main interest is to eventually simulate low-frequency turbulence (along with possible stabilization mechanisms) that saturates on much shorter time scales. In this vein, we employed the long wavelength gyrokinetic solver in the \gkeyll~code since it is capable of future higher dimensional, turbulence simulations. Three different approaches to computing the potential were used: isothermal Boltzmann electrons with ambipolar sheaths, kinetic electrons with a linearized polarization, and kinetic electrons with a nonlinear polarization. This work is unique in that, compared to other efforts, it uses realistic \wham~parameters, simulates the entire plasma, including the expanders, incorporates the effects of collisions and contains kinetic electron simulations (including a sheath model).

This manuscript also presents a novel Hamiltonian-structure-preserving force softening strategy to mitigate the small time step associated with the strong mirror force produced by HTS coils. This technique was used to accelerate Boltzmann electron simulations by a factor of 19.5 and had the added benefit of removing a numerical accumulation of $f_i$ resulting from the finite velocity space extents. It was shown that the resulting ambipolar potential with and without force softening was the same, buttressing justification for its use. Other metrics, however, like the perpendicular temperature, exhibit modest modifications that may be attributed to the use of force softening. Future work will explore these modifications further; one improvement, for example, is to include an estimate of the potential profile (e.g. obtained through other codes or semianalytic methods) in the design of the softening function $\chi$.

Other computational challenges encountered in applying the \gkeyll~continuum gyrokinetic solver to the HTS mirror environment include the long integration times to build the equilibrium potential, the high-resolution grids needed to resolve colossal changes in the distribution function across the entire machine, and the small time step that the $\omega_H$ mode also imposes. In future 1D and higher dimensional simulations, we will start from an equilibrium obtained via faster means, either \gkeyll~Boltzmann electron simulations, semi-analytic methods~\cite{Egedal2022}, or bounce-averaged Fokker-Planck codes. In that case, 1D simulations can verify that the system stays close to such an equilibrium, and higher dimensional simulations will integrate for much shorter time periods (e.g. the interchange time scale $\sim1\,\mu\mathrm{s}$). Additional algorithmic improvements are also being implemented in \gkeyll, such as nonuniform grids and additional parallelism, which we estimate can speed up the code by a factor of 10-15. Lastly, there are semi-implicit algorithms that could be explored to overcome the $\omega_H$ restriction on $\Dt$.

All in all, the Boltzmann electron and nonlinear polarization kinetic electron simulations with force softening were performed out to $t=100\,\mu\mathrm{s}\approx5.56\nu_{ee}^{-1}$. It was shown that they developed an electrostatic potential comparable to that estimated using analytic formulas in the early calculations of this phenomenon~\cite{Pastukhov1974,Cohen1978,Chernin1978}. This means that the gyrokinetic model used does evolve towards an equilibrium that confines the electrons and that, in the future, when we pursue turbulence studies about an externally computed equilibrium, the system will not deviate from such equilibrium. A caveat is that the nonlinear polarization simulation exhibited a late-stage electron heating, causing the $e\Delta\phi/T_e(z=0)$ ratio to decrease some; exploring the origin of this heating is planned for our immediate future work. Additionally, as an extra check on these two simulations, we also pursued kinetic electron simulations with a linearized polarization and no force softening. However, this simulation required a much smaller $\Dt$ and was only advanced for a third of the time of the other two. Thus far, the linearized polarization model appears to track the nonlinear polarization model, providing some partial confidence in the nonlinear polarization result. Finally, we note that beyond the computational and field model improvements discussed, the physics model may need considerable enhancements to achieve predictive capabilities. For example, an essential ingredient of \wham~dynamics is the presence of sloshing ions due to their modification of the pressure and potential profiles and their effect on plasma instabilities. Another aspect of the model that may need to be enhanced is the use of a Dougherty instead of the LRO. These are left for future work.

\begin{acknowledgments}
We thank Ian Abel and Felix Parra for sharing their insights on this topic and express our gratitude towards other members of the \gkeyll~and \wham~teams who aided this work. We used the Stellar cluster at Princeton University. M.F., A.H. and G.W.H. were supported by the Partnership for Multiscale Gyrokinetic Turbulence (MGK) and the High-Fidelity Boundary Plasma Simulation (HBPS) projects, part of the U.S. Department of Energy (DOE) Scientific Discovery Through Advanced Computing (SciDAC) program, and the DOE's ARPA-E BETHE program, via DOE contract DE-AC02-09CH11466 for the Princeton Plasma Physics Laboratory.
\end{acknowledgments}

\section*{Data Availability Statement}

Raw data were generated at PPPL/Princeton University Stellar computer cluster. The derived data suppporting this study's findings are openly available in Zenodo at \url{https://doi.org/10.5281/zenodo.7738951}.

\appendix

\section{Energy conservation of 1D gyrokinetic models} \label{sec:energyConserv}

The kinetic electron models in section~\ref{sec:gkmodels} conserve the total energy~\cite{Sugama2000} 
\begin{equation}
    W_{\text{tot}} = W_{H} - L_f = \sum_s\int\dvtz\,\GJac\BparS f_s H_s - L_f,
\end{equation}
where $\dvtz=\dz\,\dvw$ and $L_f$ is the field part of the Lagrangian. This part, and the Hamiltonian $H$, needs to be suitably defined depending on which field model is used, as shown below. Note that in both cases $L_f$ does not include the familiar term proportional to $\epsilon_0\abs{\v{E}}^2/(8\pi)$ because we order that as small in the Poisson equation and use quasineutrality instead.

\subsection{Linear polarization} \label{sec:enerConservLinear}

When using the gyrokinetic model consisting of equations~\ref{eq:gkeq1d}-\ref{eq:poissonBracket1d},~\ref{eq:poisson1dlinear}, we employ the lower order Hamiltonian $H=m_s\vpar^2/2+\mu B+q_s\phi$. However to obtain a field equation for $\phi$ and have a conserved total energy, one must employ the field Lagrangian $L_f=\sum_s\int\dz\,\GJac[m_sn_0/(2\Bp^2)]\abs{\gradperp{\phi}}^2$, such that the total conserved energy is
\begin{equation} \label{eq:wtot}
    W_{\text{tot}} = W_{H} - \int\dz\,\GJac\frac{\epsperp}{2}\abs{\gradperp{\phi}}^2,
\end{equation}
where the linear polarization model uses $\epsperp=\sum_s m_sn_0/\Bp^2$ and replaces $\gradperp{\phi}\to-\kperp\phi$ in the last term.
The time evolution of $W_H$ given by
\begin{eqnal} \label{eq:hamilEner1}
    \d{W_H}{t} 
    &= \int\dz\,\GJac\sum_s\left(\int\dvw~\BparS\pd{f_s}{t}H_s+q_sn_s\pd{\phi}{t}\right).
\end{eqnal}
The contribution from the first term can be obtained by multiplying the gyrokinetic equation~\ref{eq:gkeq1d} by $H$ and integrating over all phase space (disregarding collisions since they conserve energy~\cite{Francisquez2020})
\begin{eqnal}
    &\sum_s\int\dvtz\,\GJac\BparS\pd{f_s}{t}H_s \\
    &\quad= \sum_s\int\dvtz\left[-\pd{}{z}\left(\vpar f_s\right) + \frac{1}{m_s}\pd{H_s}{z}\pd{f_s}{\vpar}+\GJac\BparS S_s\right]H_s.
\end{eqnal}
Integrating the second term by parts leads us to
\begin{eqnal}
    &\sum_s\int\dvtz\,\GJac\BparS\pd{f_s}{t}H_s
    = \sum_s\int\dvtz\left[-\pd{}{z}\left(\vpar f_s\right) + \GJac\BparS S_s\right]H_s \\
    &\quad+ \sum_s\frac{2\pi}{m_s^2}\int\dz\,\dmu\pd{H_s}{z}f_sH_s\Big|^{\vparmax}_{\vparmin} - \sum_s\int\dvtz\,\pd{H_s}{z}\frac{f_s}{m_s}\pd{H_s}{\vpar}.
\end{eqnal}
The second term on the right side vanishes due to zero-flux BCs along $\vpar$, and the expression can be consolidated into
\begin{eqnal} \label{eq:hamilEner2}
    &\sum_s\int\dvtz\,\GJac\BparS\pd{f_s}{t}H_s
    = \sum_s\int\dvtz\left[-\vpar\pd{}{z}\left(f_sH_s\right) + \GJac\BparS S_s H_s\right] \\
    &\quad= -\sum_s\int\dvw\,\vpar f_sH_s\Big|^{\zmax}_{\zmin} + \sum_s\int\dvtz\,\GJac\BparS S_s H_s.
\end{eqnal}
With the use of equations~\ref{eq:hamilEner1} and~\ref{eq:hamilEner2}, the time derivative of equation~\ref{eq:wtot} yields the following total energy time rate of change
\begin{eqnal}
    \d{W_{\text{tot}}}{t} &= -\sum_s\int\dvw\,\vpar f_sH_s\Big|^{\zmax}_{\zmin} + \sum_s\int\dvtz\,\GJac\BparS S_s H_s \\
    &\quad+ \int\dz\,\GJac\sum_sq_sn_s\pd{\phi}{t} - \int\dz\,\GJac\epsperp\kperp {\phi}\pd{}{t}\kperp{\phi}.
\end{eqnal}
Now replace $\sum_sq_sn_s$ using the field equation~\ref{eq:poisson1dlinear} causing the last two terms to cancel each other, and obtain
\begin{eqnal} \label{eq:dWdtlin}
    \d{W_{\text{tot}}}{t}
    &= -\sum_s\int\dvw\,\vpar f_sH_s\Big|^{\zmax}_{\zmin} + \sum_s\int\dvtz\,\GJac\BparS S_s H_s.
\end{eqnal}
Therefore, any change in the total energy of the system is only caused by the energy fluxes through the sheath entrance (first term in equation~\ref{eq:dWdtlin}) or by the sources (second term in equation~\ref{eq:dWdtlin}).

\subsection{Nonlinear polarization} \label{sec:enerConservNonlinear}

The energy theorem for the model with a nonlinear polarization density proceeds in a similar way, except that in this case $L_f=0$ and we must use the higer-order Hamiltonian $H=m_s\vpar^2/2+\mu B+q_s\phi-m_s\left(\kperp\phi/\Bp\right)^2/2$. Therefore the time rate of change of the total energy is therefore
\begin{eqnal} \label{eq:enerNL1}
    \d{W_H}{t} 
    &= \int\hspace{-2pt}\dz\,\GJac\sum_s\left[\int\hspace{-2pt}\dvw\,\BparS\pd{f_s}{t}H_s+n_s\left(q_s-m_s\frac{\kperp^2\phi}{\Bp^2}\right)\pd{\phi}{t}\right].
\end{eqnal}
Similar arguments as in the previous subsection reduce the first term on the right to the right side of equation~\ref{eq:hamilEner2}, such that our total energy rate of change in this case is
\begin{eqnal}
    \d{W_{\text{tot}}}{t}
    &= -\sum_s\int\dvw\,\vpar f_sH_s\Big|^{\zmax}_{\zmin} + \sum_s\int\dvtz\,\GJac\BparS S_s H_s \\
    &\quad+ \int\dz\,\GJac\sum_sn_s\left(q_s-m_s\frac{\kperp^2\phi}{\Bp^2}\right)\pd{\phi}{t}. \\
    %
    %
\end{eqnal}
Using the field equation~\ref{eq:poisson1dnonlinear} to replace $\sum_sq_sn_s$ in the term on the right causes the round bracket to vanish, and we arrive at equation~\ref{eq:dWdtlin} again, i.e. energy change is only caused by boundary fluxes or sources.

\section{Electrostatic shear Alfv\'en or $\omega_H$ modes} \label{sec:omegaH}

\subsection{Linear polarization} \label{sec:omegaHlin}

In the electrostatic limit the equations in our model contain a high frequency wave mode that is usually named the electrostatic shear Alfv\'en or $\omega_H$ mode. In order to get an estimate of its frequency, we consider the electrostatic, collisionless slab limit of equation~\ref{eq:gkeq1d} without sources and gradients in $B$. That is:
\begin{equation}
\pd{f_s}{t} + \pd{}{z}\left(\vpar f_s\right) - \pd{}{\vpar}\left(\frac{q_s}{m_s}\pd{\phi}{z} f_s\right) = 0.
\end{equation}
By linearizing this equation about an instantaneous Maxwellian equilibrium ($f_{s0}$) with zero electrostatic potential ($\phi_0=0$), we obtain an equation for the perturbation $f_{s1}$:
\begin{equation}
\pd{f_{s1}}{t} + \pd{}{z}\left(\vpar f_{s1}\right) - \pd{}{\vpar}\left(\frac{q_s}{m_s}\pd{\phi_1}{z} f_{s0}\right) = 0.
\end{equation}
We assumed that the spatial variation in the Maxwellian $f_{s0}$ with density $n_{s0}(z,t)$ and temperature $T_{s0}$ is very weak compared to the strong inhomogeneity induced by this mode, i.e. $\kpar n_{s0}(z,t)/\abs{\partial_zn_{s0}(z,t)} \gg1$, where $\kpar$ is a mode wave number along the field line. The $\vpar$ derivative is trivially calculated given the Maxwellian form of $f_{s0}$, yielding
\begin{equation} \label{eq:gkeq1dlin}
\pd{f_{s1}}{t} + \pd{}{z}\left(\vpar f_{s1}\right) + \frac{q_s}{m_s}\pd{\phi_1}{z}\frac{\vpar}{v_{ts0}^2}f_{s0} = 0.
\end{equation}
We can now employ the ansatz $f_{s1}=\hat{f}_{s1}\exp\left(-i\omega t+i\kpar z\right)$ and $\phi_{1}=\hat{\phi}_{1}\exp\left(-i\omega t+i\kpar z\right)$, to simplify equation~\ref{eq:gkeq1dlin} to
\begin{equation} \label{eq:gkeq1dlink}
-i\omega f_{s1} + i\kpar\vpar f_{s1} + \frac{q_s}{m_s}i\kpar\phi_1\frac{\vpar}{v_{ts0}^2}f_{s0} = 0.
\end{equation}

In order to make further progress the relationship between $f_{s1}$ and $\phi_1$ must be exploited. We assume electrons and ions are in quasineutral equilibrium. In that case, the field equation with a linearized polarization simply states that
\begin{equation} \label{eq:poissoneq1dlin}
    \kperp^2\sum_s\frac{n_0m_s}{B_p^2}\phi_1 = \sum_s q_sn_{s1}.
\end{equation}
We further assume (one may check a posteriori) that the ions are too heavy and slow to respond to this fast electrostatic wave so that we can assume that they are static. Thus only the electron evolution need be considered with equations~\ref{eq:gkeq1dlink}-\ref{eq:poissoneq1dlin}, which may be combined to yield
\begin{equation} 
    \left(\int\dvw\,\BparS \frac{q_e^2}{m_ev_{te0}^2}\frac{\kpar\vpar}{\omega-\kpar\vpar}f_{e0} - \kperp^2\frac{n_0m_i}{B_p^2}\right)\phi_1 = 0.
\end{equation}
For this electrostatic wave to exist the term inside the brackets must vanish. Furthermore, we may perform the $\mu$ integral since the dependence on this variable is only through the Maxwellian equilibrium $f_{e0}$:
\begin{equation} 
    \frac{q_e^2}{m_ev_{te0}^2}\int\dvpar\,\frac{\kpar\vpar}{\omega-\kpar\vpar}f_{e0\parallel} - \kperp^2\frac{n_0m_i}{B_p^2} = 0,
\end{equation}
and $f_{e0\parallel}$ is a Maxwellian in $\vpar$ with $n_{e0}(z,t)$ and $T_{e0}$ density and temperature, respectively. We can write this in terms of the plasma dispersion function
\begin{equation} 
    \kperp^2\frac{m_i}{B_p^2} + \frac{q_e^2}{m_ev_{te0}^2}\frac{n_{e0}(z,t)}{n_0}\left[1+\frac{\omega}{\sqrt{2}v_{te0}\kpar} Z\left(\frac{\omega}{\sqrt{2}v_{te0}\kpar}\right)\right] = 0.
\end{equation}
Expanding the plasma dispersion function to first order in $\omega/(\sqrt{2}v_{te0}\kpar)\gg1$, we obtain
\begin{equation} 
    \omega = \omega_H = \frac{v_{te0}\kpar}{\kperp\rho_{s0}}\sqrt{\frac{n_{e0}(z,t)}{n_0}}.
\end{equation}

\subsection{Nonlinear polarization} \label{sec:omegaHnonlin}

In this case we make a similar analysis as in the previous section, with the modification that we allow incorporate the instantaneous spatial variation of the density profiles into the polarization density:
\begin{equation} \label{eq:poissoneq1dnonlin}
    \kperp^2\sum_s\frac{n_{s0}(z,t)m_s}{B_p^2}\phi_1 = \sum_s q_sn_{s1}.
\end{equation}
Again we highlight that the spatial and temporal variation implied by the arguments of $n_{s0}(z,t)$ is much weaker than that of the $\omega_H$ mode, allowing us to perform a strictly local linear analysis about the $n_{s0}(z,t)$ background density. The rest of the derivation proceeds as before, leading to the dispersion relation
\begin{equation} 
    \kperp^2\frac{m_i}{B_p^2} + \frac{q_e^2}{m_ev_{te0}^2}\frac{n_{e0}(z,t)}{n_{i0}(z,t)}\left[1+\frac{\omega}{\sqrt{2}v_{te0}\kpar} Z\left(\frac{\omega}{\sqrt{2}v_{te0}\kpar}\right)\right] = 0,
\end{equation}
whose high frequency limit is
\begin{equation} 
    \omega = \omega_H = \frac{v_{te0}\kpar}{\kperp\rho_{s0}}\sqrt{\frac{n_{e0}(z,t)}{n_{i0}(z,t)}}.
\end{equation}

\section{Resolution scan} \label{sec:resconv}

In order to arrive at the lowest possible resolution that would produce acceptable results we tried several resolutions during the course of this work. We note that when we first attempted kinetic electron simulations, the simulations were not stable if $\Nz<256$. Catastrophic oscillations appeared near the mirror throats when we used fewer than 256 cells. This is presumably due to large gradients in $f_s$ that occur with a high mirror ratio, leading to regions of $f_s<0$. \gkeyll~can tolerate some $f_s$ negativity but, if it gets too strong, unphysical instabilities can arise and cause the simulation to diverge. A positivity-preserving algorithm, mentioned in section~\ref{sec:compfuture}, would help prevent such instabilities and allow the use of coarser grids to produce useful results. It is possible that such instabilities were exacerbated by incorrect velocity resolutions used initially (more on this below), and that once a more appropriate $\Nvpar\times\Nmu$ was selected a lower $\Nz$ could be used. We have not yet explored that possibility, but will in the future. For this manuscript the number of cells along $z$ was later increased to $288$ because the simulations were carried out in nodes with 96 2.9 GHz Intel Cascade Lake cores, so this allowed us to parallelize across 3 nodes in a balanced manner with marginal computational cost.

\begin{figure}
\includegraphics[width=0.48\textwidth]{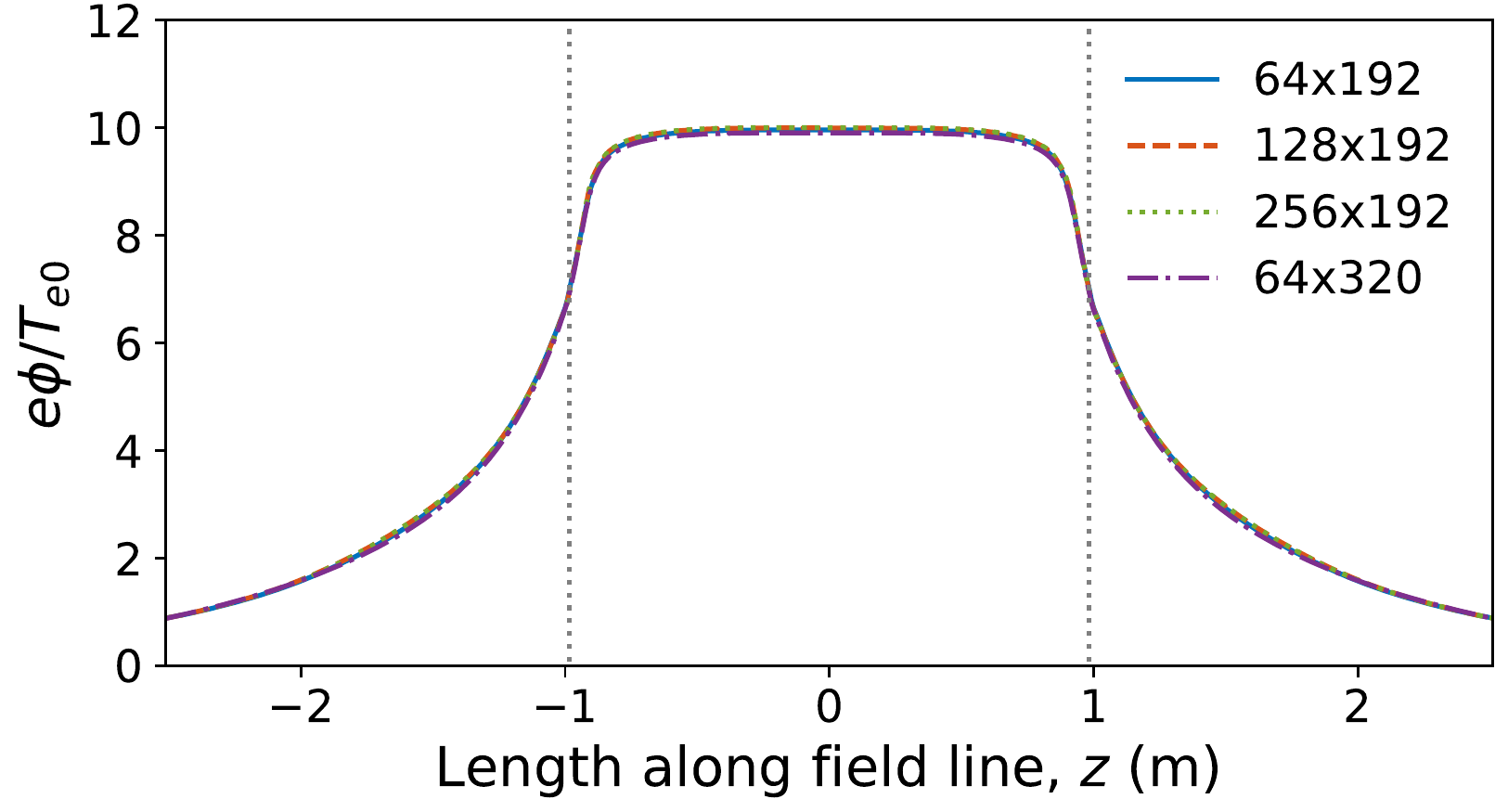}
\caption{\label{fig:phiresScanAdi} Electrostatic potential profile at $t=10.5\,\mu\mathrm{s}$ in Boltzmann electron simulations with $\Nz=288$ and several $\Nvpar\times\Nmu$ resolutions.}
\end{figure} 

\begin{figure}
\includegraphics[width=0.48\textwidth]{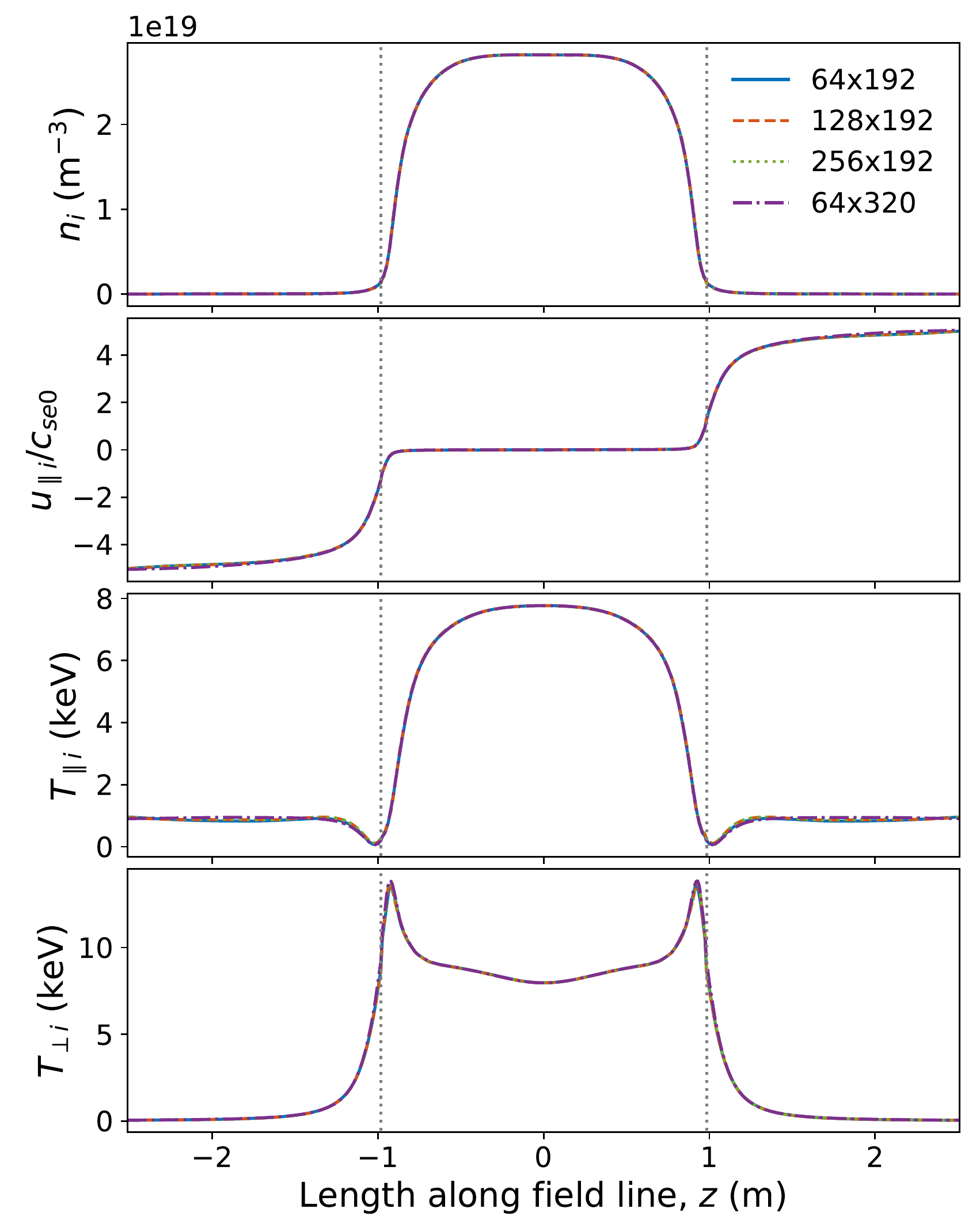}
\caption{\label{fig:ionMomresScanAdi} Ion number density ($n_i$), mean parallel speed ($\upari$) normalized to $c_{se0}=\sqrt{T_{e0}/m_i}$, and parallel and perpendicular temperatures ($\Tpari,\,\Tperpi$) at $t=10.5\,\mu\mathrm{s}$ in Boltzmann electron simulations with $\Nz=288$ and several $\Nvpar\times\Nmu$ resolutions.}
\end{figure}

After establishing the position-space resolution, we performed a limited scan of the velocity space resolution. These simulations were only run for a few microseconds (see figure captions) given their considerable cost, especially at higher resolution. Four Boltzmann electron simulations with the resolutions $\Nvpar\times\Nmu=\{64\times192,128\times192,256\times192,64\times320\}$ were performed up to $10.5\,\mu\mathrm{s}$. As can be seen from figure~\ref{fig:phiresScanAdi} the electrostatic potential profile did not change for any of these resolutions. The velocity moments of the ion distribution ($n_i$, $\upari$, $\Tpari$ and $\Tperpi$) also were relatively unaffected by the various resolutions employed here (figure~\ref{fig:ionMomresScanAdi}). We should note that these are ion profiles whose steady state is likely not reached at this point, but we had to make an assessment of the resolution requirements with the simulation time that our resources allowed. 

\begin{figure}
\includegraphics[width=0.48\textwidth]{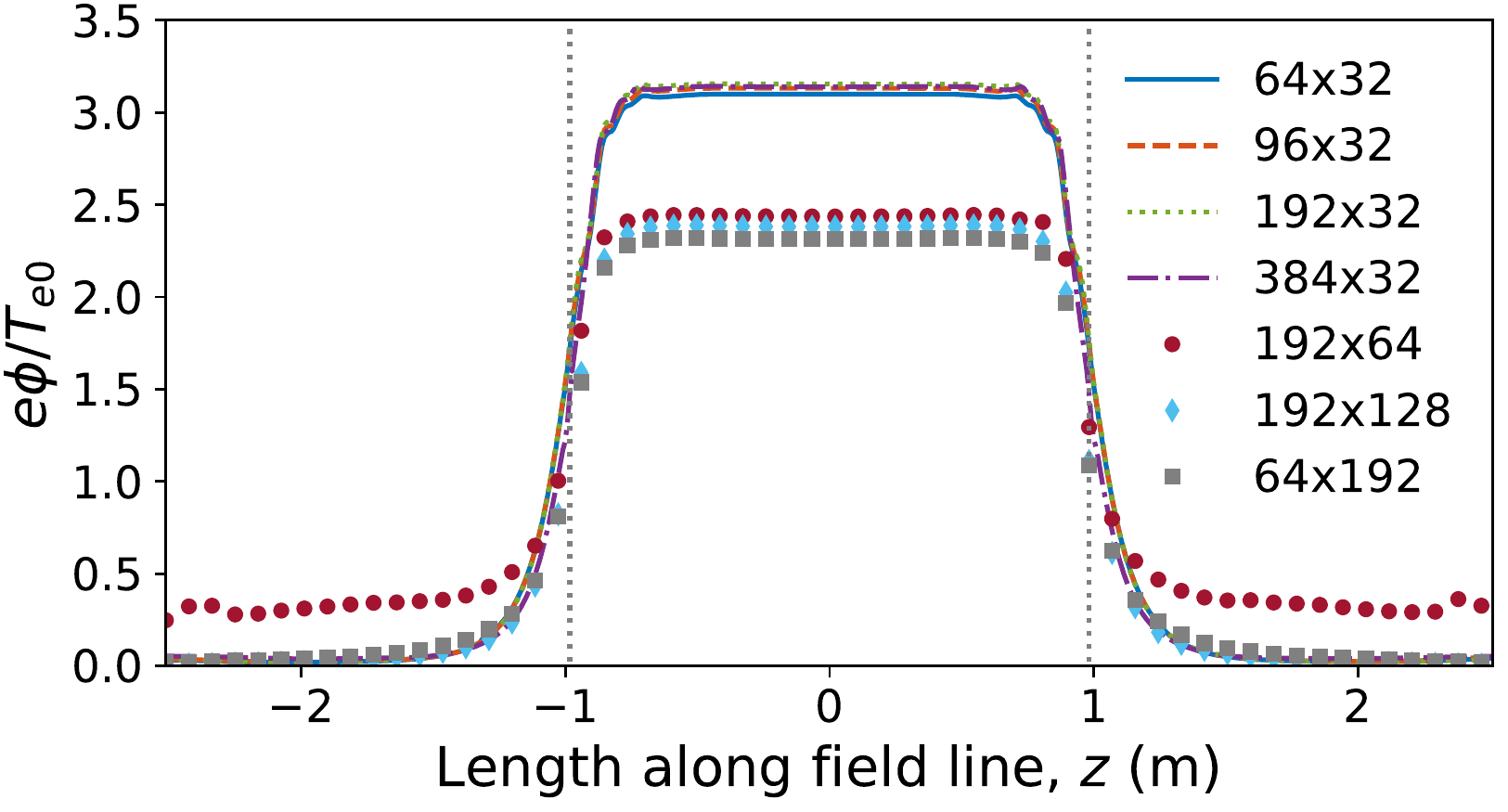}
\caption{\label{fig:phiresScan} Electrostatic potential profile at $t=6\,\mu\mathrm{s}$ in linearized polarization kinetic electron simulations with $\Nz=288$ and several $\Nvpar\times\Nmu$ resolutions.}
\end{figure}

Future Boltzmann electron simulations could possibly be run longer (incorporating computational improvements mentioned in section~\ref{sec:compfuture}) or with lower resolution. For now we have opted to carry on with $\Nz\times\Nvpar\times\Nmu=288\times64\times192$ as our baseline, especially given some of the findings when scanning resolution with kinetic electrons. Our kinetic electron resolution scan used the linearized polarization model with $\kperp\rhos=0.01$ (and no force softening), and were run until $t=6\,\mu\mathrm{s}$. Note that this is a more demanding $\kperp\rhos$ choice than the $\kperp\rhos=0.06$ estimated for \wham~with $\Bp=0.86$~(see section~\ref{sec:resultsKinElc}) due to the $\omega_H$ mode; nonlinear polarization with a higher $\kperp\rhos$ could be used in the future to carry out these convergence tests and other simulations much more quickly. Once again we kept the $z$ resolution fixed at 288 cells, and proceeded to vary $\Nvpar\times\Nmu$. This scan produced the electrostatic potential profiles in figure~\ref{fig:phiresScan}; it shows that regardless of $\Nvpar=64-384$, if the $\mu$ resolution is kept low (32 cells) a considerably higher central potential is produced. Only when $\Nmu\geq64$ do we see $\phi(z=0)$ begin to asymptote, but approaching convergence in the expander seems to further necessitate $\Nmu\geq128$. An justification of such high resolution $\mu$ grids was offered in section~\ref{sec:comp}, and ideas to improve on this with variable $\mu$ grids were discussed in section~\ref{sec:compfuture}.

\begin{figure}
\includegraphics[width=0.48\textwidth]{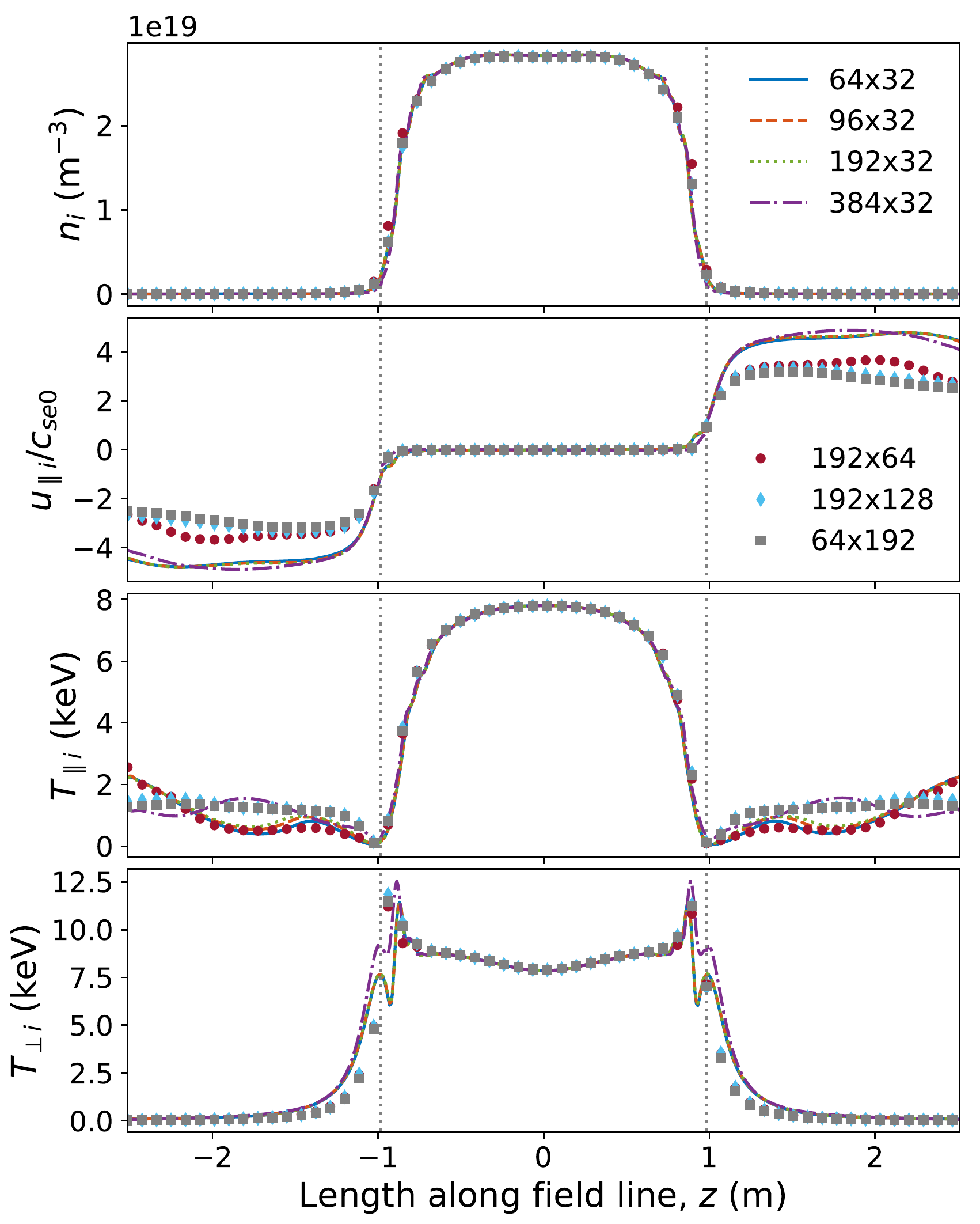}
\caption{\label{fig:ionMomresScan} Ion number density ($n_i$), mean parallel speed ($\upari$) normalized to $c_{se0}=\sqrt{T_{e0}/m_i}$, and parallel and perpendicular temperatures ($\Tpari,\,\Tperpi$) at $t=6\,\mu\mathrm{s}$ in linearized polarization kinetic electron simulations with $\Nz=288$ and several $\Nvpar\times\Nmu$ resolutions.}
\end{figure} 

\begin{figure}
\includegraphics[width=0.48\textwidth]{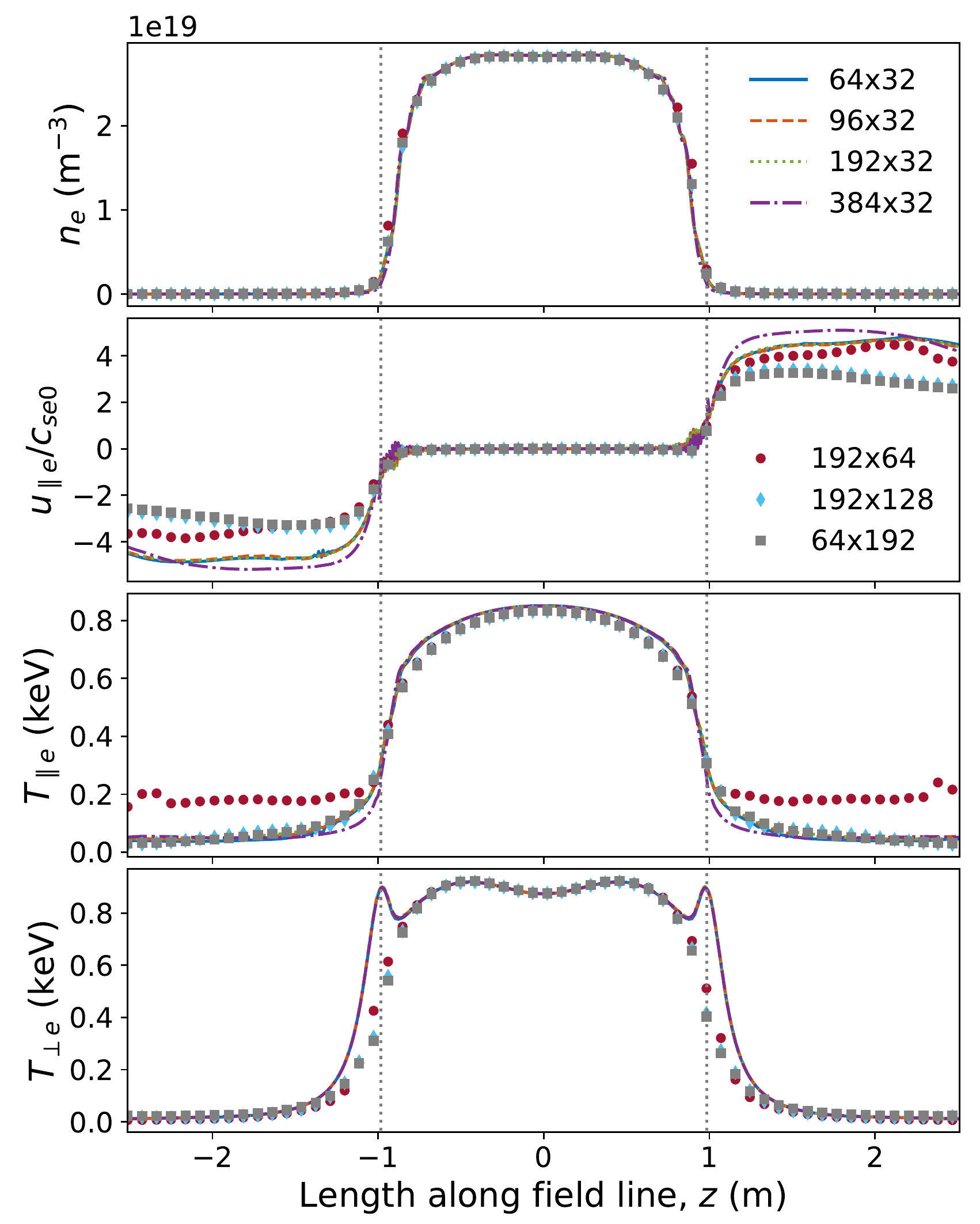}
\caption{\label{fig:elcMomresScan} Electron number density ($n_e$), mean parallel speed ($\upare$) normalized to $c_{se0}=\sqrt{T_{e0}/m_i}$, and parallel and perpendicular temperatures ($\Tpare,\,\Tperpe$) at $t=6\,\mu\mathrm{s}$ in linearized polarization kinetic electron simulations with $\Nz=288$ and several $\Nvpar\times\Nmu$ resolutions.}
\end{figure}

The ion velocity moments were also examined for each of these simulations (figure~\ref{fig:ionMomresScan}). We first notice that unlike the Boltzmann electron simulations there is more variability with resolution here, although it is possible that is simply caused by these snapshots being taken earlier in time. As with the potential profiles, we see that some features take a noticeable leap when going from $\Nmu=32$ to $\Nmu=64$; see for example the ion parallel speed in figure~\ref{fig:ionMomresScan}(b). An additional but smaller change is obtained when $\mu$ resolution is further increased to $128$ cells. Similar trends are seen in the electron velocity moments (figure~\ref{fig:elcMomresScan}), which led us to conclude that a $288\times64\times192$ grid was the most suitable for this study. Note that the velocity grids do not have to be the same for both species, and in the future it is possible that the cost can be further reduced by using a coarser mesh for ions.

\nocite{*}
\bibliography{main}

\end{document}